\documentclass[preprint2]{aastex}

\shorttitle{The multiple periods of DI~Cru}
\shortauthors{Oliveira et al.}

\begin{document}

\title{The multiple spectroscopic and photometric periods \\ of \objectname{DI~Cru} $\equiv$ \objectname{WR~46}
\thanks{Based on observations made at Observat\'orio do Pico dos Dias/LNA, Brazil, and at
	the 1.5m ESO telescope at La Silla, Chile.}
       }

\author{Alexandre S. Oliveira\altaffilmark{2}, J. E. Steiner\altaffilmark{2} and M. P. Diaz\altaffilmark{2}}
\altaffiltext{2}{Instituto de Astronomia, Geof\'{\i}sica e Ci\^encias Atmosf\'ericas, Universidade de S\~ao Paulo, CP 9638, 01065-970
               S\~ao Paulo, Brazil}
\email{alex@astro.iag.usp.br,steiner@astro.iag.usp.br,marcos@astro.iag.usp.br}

\begin{abstract}

In an effort to determine the orbital period of the enigmatic star \objectname{DI~Cru} $\equiv$ \objectname{WR~46}
$\equiv$ \objectname{HD~104994}, we made 
photometric and spectroscopic observations of this object between 1996 and 2002. Both photometric 
and spectroscopic characteristics are quite complex. The star is highly variable on short (few hours) as 
well as on long (few months) time-scales.

The optical spectrum is rich in strong emission lines of high ionization species 
such as \ion{He}{2}, \ion{N}{4}, \ion{N}{5} and \ion{O}{6}. Weak emission 
of \ion{C}{3} and H$\beta$ is also present. Emission lines have been compiled and identified from the ultraviolet to the infrared. In
the UV, emission of \ion{O}{5} and \ion{N}{4} is also observed, together with very weak emission of \ion{C}{4}. The 
\ion{N}{5} 4603+19{\AA}/\ion{He}{2} 4686{\AA} line ratios vary by a significant amount from night to night. Temporal Variance 
Spectrum -- TVS -- analysis shows that the \ion{He}{2} 4686{\AA} line has P Cyg-like variable absorption, while \ion{N}{5} 4603/19{\AA}
 lines have a strong and broad variable component due to the continuum fluorescence from a source (stellar 
atmosphere/optically thick wind) of  variable temperature. We also show that the object has variable degree of 
ionization, probably caused by wind density variation.

The star presents multiple periods in radial velocity and photometry. We derived, from our data, a main 
radial velocity period  of 0.3319 d with an amplitude of $K = 58$ km~s$^{-1}$. This period is similar to the value found by 
\citet{marche2}. When at intermediate brightness level, this period is also seen in the photometric
 measurements. When the star is at bright phase, the photometric variations do not present the same period. 
 Photometric periods ranging from 0.154 to 0.378 d are present, consistent with observations reported by other authors. 
 Besides the 0.3319~d period other spectroscopic periods are also seen. On distinct epochs, the periods are different;  
 \citet{marche2} interpreted the 0.3319 d period as the orbital one. Although we do not discard this possibility, the true binary 
 nature (e.g. long term coherence or detection of a secondary star) has not yet been demonstrated.
 
 DI~Cru is a Population I WR object. Given the similarities (e.g. multiple periods likely due to non-radial oscillations),
 it could be interpreted as a luminous counterpart of the qWR star HD~45166.

\end{abstract}


\section{Introduction}

DI~Cru $\equiv$ WR~46 $\equiv$ HD~104994 is a Wolf Rayet star that has been classified as a peculiar early WN, weak lined object. In 
The VIIth Catalogue of Galactic Wolf-Rayet Stars \citep{huc01} it was classified as WN3~pec, where the ''pec'' 
suffix indicates the presence of strong \ion{O}{6} 3811/34{\AA} emission lines. There are only three WR stars of the nitrogen 
sequence in this catalog that display strong \ion{O}{6} emission lines. The other stars are \objectname{WR~109} (also known as 
\objectname{V617~Sgr} \citep{steiner99} and \objectname{WR~48c}, also known as \objectname{WX~Cen}
\citep{dia95}. Both of them have well defined short orbital periods. \objectname{WR~156} is also reported in the Catalog as having 
\ion{O}{6} emission lines. However, the \ion{N}{5} 4945{\AA} line is not present in the spectrum published by \citet{marche3} and 
\ion{O}{6} lines are not visible in the spectrum published by \citet{hama}. DI~Cru was classified by \citet{smit}
-- SSM96 -- in their three-dimensional classification system as WN3b pec. The letter ''b'' stands for 
FWHM(\ion{He}{2} 4686{\AA) $> 30$ {\AA} or 1920 km~s$^{-1}$.

The Einstein Observatory IPC detected DI~Cru as a source with a derived X-ray luminosity $L_X(0.2-4~ KeV) \sim 3 \times 10^{33}$
erg~s$^{-1}$ with an estimated distance of 8.7 kpc \citep{pol}. It was also detected by the ROSAT PSPC instrument \citep{wes}. 
\citet{huc88} list a distance of 3.44 kpc and an absolute magnitude of $M_V = -2.8$. In their analysis of Wolf-Rayet stars 
distribution in the Galaxy, on the basis of their distance determination by spectroscopic parallaxes, \citet{con} estimated 
a distance of 6.3 kpc, assuming $M_V=-3.8$ and $(b-v)_0 = -0.20$. These values were derived under the hypothesis 
that the star has standard WN3 properties. \citet{vee00} derived a lower limit of 1.0 kpc based on the upper limits to the radio fluxes at 3 and 6 cm and on 
an adopted mass-loss rate of $ log \dot{M}~(M_{\sun}~ yr^{-1}) = -5.2$. In addition to this, \citet{tov}  place the system in an OB
 association with 11 other stars at a distance of 4.0 kpc.

\citet{mon} found photometric variability with 0.075 mag amplitude in the $V$ filter and time-scale of about 0.125 d. In a 
follow-up photometric study, \citet{gen} suggest a binary nature with an orbital period of 0.2824 d and an asymmetric 
0.15 mag amplitude light curve. These authors also suggest that the emission lines are emitted by a circumstellar envelope.

\citet{cro} - CSH95 - presented ultraviolet and optical observations of DI~Cru and made a detailed wind 
model analysis of this object in the context of weak-lined WNE stars. They concluded that the star is well modeled as a population 
I object with strong wind. They also concluded, however, that its location in the HR diagram and its CNO abundances are quite 
anomalous and are not predicted by theoretical model evolution for WR stars. These authors found that oxygen is overabundant 
by a factor of 2 (O/N~$\sim 0.1$) and that the upper limit for carbon is C/O~$< 0.3$. An independent determination of the distance
 to the star on the basis of the galactic rotation curve was also made by these authors, who found a distance $d=4.0 \pm~1.5$ kpc. The
derived  luminosity is $log~L/L_{\sun}= 5.53$, the 
  mass, $M=14~\pm~1$~M$_{\sun}$ and $log~\dot{M}/M_{\sun}~yr^{-1}=-5.20$.

A preliminary radial velocity curve of the emission lines of DI~Cru was published by \citet{vee95} and showed an amplitude of 
about 100~km~s$^{-1}$ with the 0.28 d period. The binary nature of the system seemed to be confirmed. However, from analysis of the radial
 velocity of the \ion{N}{5}~4603/19{\AA} and the \ion{He}{2} 4686{\AA} lines, \citet{nie} determined a period of 0.311274 d, quite distinct from the 
0.28 d period found before. These authors suggest DI~Cru to be an evolved binary system containing an accretion disc. Based on polarimetric observations 
 they estimate a distance of about 2 kpc and an absolute magnitude of $M_V=-1.6$. \citet{gen} and also \citet{vee95} 
 proposed that the star is a binary system in which the low mass object is a white dwarf -- a configuration difficult to explain considering 
 the theory of binary evolution.

\citet{marche1} reported the discovery, using the Hipparcos satellite, of long term photometric variations with approximately 
0.2 mag amplitude ($V_j$). Long term variations in the equivalent width of the \ion{He}{2} 5411{\AA} emission line correlated with these photometric 
variations were also reported \citep{vee99}.
\citet{marche2} - Mar00 - presented extensive spectroscopic and photometric monitoring of DI~Cru and showed that the star
 reveals periodic variations with a period of $0.329~\pm~0.013$ d. Although these authors interpret the system as a classical population 
 I WR star in a binary system, they found that the radial velocity modulation disappears from time to time, establishing a new puzzle.

Recently, \citet{vee02a,vee02b,vee02c} -- Vee02a,b,c -- in a series of papers, made a detailed analysis of photometric and spectroscopic data. 
They found that the star has presented distinct periods and interpreted this as evidence for either multiple non-radial pulsation
periods or gradual change 
of the underlying clock rate. They discussed a variety of scenarios to explain the set of observations and ruled out the model of single star 
rotation as the main cause of photometric and spectroscopic variability. Between the two remaining possibilities -- non-radial 
pulsation or binary signature -- they favored the former one but concluded that the enigma of DI~Cru has not yet been solved, leaving 
room for further investigation.

\citet{steiner98} included this star in a group of 4 objects they called the V Sagittae stars. This group of objects is composed 
of galactic binary stars that share many photometric and spectroscopic observational properties that are not found among the 
canonical cataclysmic variables or WR stars. The stars of this class are spectroscopically characterized by the simultaneous 
presence of emission lines of \ion{O}{6} and \ion{N}{5} and by the strength of the \ion{He}{2} 4686{\AA} emission line 
relative to H$\beta$ usually larger than 2. There is also indication of strong wind in all the systems, as can be concluded from the clear P Cygni 
profiles, seen in all objects (see also \citet{steiner99} and \citet{dia99}). \ion{He}{1} lines are very weak or most frequently absent. 
No evidence of atmospheric absorption features from the secondary star has been published so far. The other three members of this 
group are:  \objectname{V~Sge} \citep{her}, V617~Sgr = WR~109 \citep{steiner99,cie} and WX~Cen = WR~48c 
\citep{dia95}. The orbital periods of these stars are 12 hr, 5 hr and 10 hr, respectively. One possible interpretation 
of these stars is that they are the galactic counterpart of the Close Binary Supersoft X-ray Sources (CBSS) seen in the Magellanic 
Clouds and in M31 \citep{steiner98}. The most popular model to explain the properties of the CBSS is that of hydrostatic 
nuclear burning on the surface of a white dwarf  \citep{heu}. This stable nuclear burning can occur when the mass 
transfer ratio is very high ($10^{-7}~ M_{\sun}~yr^{-1}$), a situation that occurs in systems with mass ratios inverted ($q=M_2/M_1 <1$) when 
compared to those usually found in cataclysmic variables \citep{kah}.

The suggestion that DI~Cru is a V Sagittae star has been criticized by Mar00 and by Vee02a,b. Their main arguments are: 
a) unlike CBSS/V~Sge stars, DI~Cru has a strong wind and b) the population I model calculated by \citet{cro} describes well 
the line profiles, and the derived properties are similar to other weak lined WNE stars. As we will show in Section~\ref{natu}, we argue that 
DI Cru is a population I WR and not a V Sge star. However, the first of the two arguments above is not correct since both classes
present strong wind. It is curious that these two groups of objects are members of completely distinct stellar paradigms and, yet,
present similar spectral properties. WR stars are post main sequence phases of very massive ($\sim 60$ M$_{\sun}$) stars
-- a young population. Currently they are in central helium (N sequence) or carbon (C sequence) burning phases.
In contrast, V Sge stars are an evolutionary phase of a binary system containing a white dwarf -- an intermediate mass, 
old population star.

In the present work we show the results of the analysis of photometric and spectroscopic data on DI~Cru obtained 
during the years 1996 -- 2002. In Section 2 we describe the observations and data reduction procedures, while the optical 
spectrum is presented and discussed in Section 3. Sections 4 and 5 present the search for periodicities in this system,
while Sections 6 and 7 present general discussions and the conclusions.

\section{Observations and data reduction}  \label{obs}

\subsection{Photometric observations}

The photometric data on DI~Cru were obtained between 1996 and 1999, in a total of 45
nights, using the 60 cm Boller \& Chivens telescope of the University of S\~ao Paulo and
the 60 cm Zeiss-Jena telescope, both at the Laborat\'orio Nacional de Astrof{\'{\i}}sica in
Itajub\'a, Brazil. In these 4 years, 4 different CCDs were used to carry out the
observations. The images were obtained through the Johnson $V$
band or in white light and cover fields between 2.7\arcmin\, x 3.6\arcmin\, and 7.9\arcmin\, x 11.9\arcmin, which contain DI~Cru and
the same four comparison stars always used in the differential photometry.

Table~\ref{jophoto} gives the journal of photometric observations.  Bias
and flatfield images were obtained to correct for undesirable instrumental
signatures. Most of the images present an overscan region used for correction offsets in the mean bias level. The data reduction
was performed in the standard way, using the IRAF \footnote{ IRAF is distributed by the National Optical Astronomy Observatories,
    which are operated by the Association of Universities for Research
    in Astronomy, Inc., under cooperative agreement with the National
    Science Foundation.} routines. Differential aperture
photometry was executed using the DAOPHOT II routines package. 
 We applied, to these data, period search routines based on Fourier
analysis with cleaning of spectral windows effects (CLEAN) \citep{roberts},
Phase Dispersion Minimization method \citep{stell} and also
Lomb-Scargle method \citep{sca}.

\placetable{jophoto}

\subsection{Spectroscopic observations}

The spectroscopic observations were carried out with the 1.6~m telescope and
the Boller \& Chivens Cassegrain and Coud\'e spectrographs. At the Cassegrain spectrograph,
we used a 1200 l/mm dispersion grating to obtain 3 spectra with $\sim$2{\AA} resolution (FWHM).
A total of 124 Coud\'e spectra were obtained with the 600 l/mm and 1800 l/mm gratings, resulting in
$\sim$0.7{\AA} and $\sim$0.2{\AA} FWHM resolution, respectively. 
Table~\ref{jospect} presents the journal of spectroscopic
observations. We obtained several bias and dome flatfield exposures to correct for the
readout pattern and sensitivity variations on the detectors. Dark current correction was not necessary.
The slit width was sized to the seeing conditions at the time of observation. We took exposures
of calibration lamps
between each star exposure to obtain the pixel-wavelength
transformation. The solutions obtained were interpolated to the individual star
exposures. The image reductions, spectra extraction and wavelength calibration were
attained with IRAF standard routines.

An additional spectroscopic observational mission was made with the FEROS -- Fiber-fed Extended Bench Optical Spectrograph -- 
 \citep{kauf}) at the 1.52~m telescope of ESO (European Southern Observatory) in La Silla, Chile. 
The FEROS spectrograph uses a bench mounted Echelle grating with reception fibers in the Cassegrain
 focus. It supplies a resolution of R~=~48000, corresponding to 2.2 pixels of 15 micrometers, and spectral coverage from 3600{\AA} 
 to 9200{\AA}. A completely automatic online reduction system is available and was adopted by us. Readout time was approximately 7 minutes. 

\placetable{jospect}

\section{The optical spectrum} \label{specsect}

\subsection{Line identification}

The FEROS spectrum, with its high resolution and wide spectral coverage from 3800{\AA} to 8800{\AA}, gives us the opportunity to 
identify the emission as well as the absorption features with high accuracy over a wide spectral range.
The spectrum of DI~Cru is quite rich in emission lines, 
mostly from high ionization species such as \ion{He}{2}, \ion{N}{4}, \ion{N}{5} and \ion{O}{6} (see Table~\ref{lineprop}). 

\placetable{lineprop}

Figures~\ref{tvs}, \ref{spec2} and \ref{spec3} show our spectra obtained at LNA. 
These figures illustrate quite well the fact that the \ion{O}{6} 3811/34{\AA} and 
\ion{N}{5} 4945{\AA} are much narrower than \ion{He}{2} 4686{\AA}, while \ion{N}{5} 4603{\AA} 
has a narrow and a broad component.

\placefigure{tvs}

\begin{figure*}
      \resizebox{\hsize}{!}{\plotone{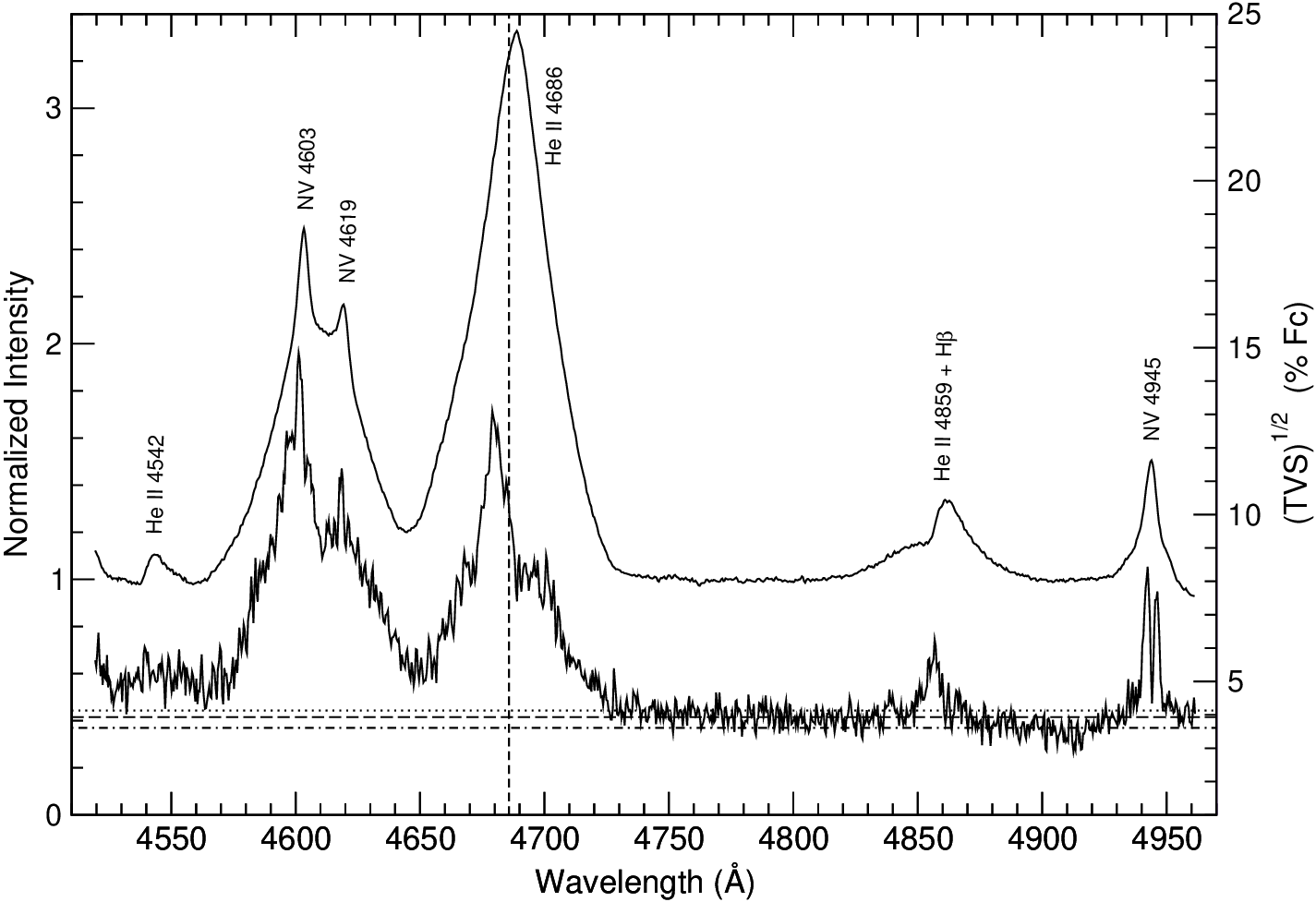}}
      \caption{Average Coud\'e spectrum (above) and TVS (below) calculated from the LNA data, 
      	both centered at 4750{\AA}. The vertical dashed line represents the \ion{He}{2} 4686{\AA}
      	rest wavelength. The TVS statistical threshold significance for p=1\%, 5\% and 30\% are represented
	by the dotted, dashed and dot-dashed horizontal lines, respectively.  \label{tvs} }
\end{figure*}

\placefigure{spec2}

\begin{figure*}
      \resizebox{\hsize}{!}{\plotone{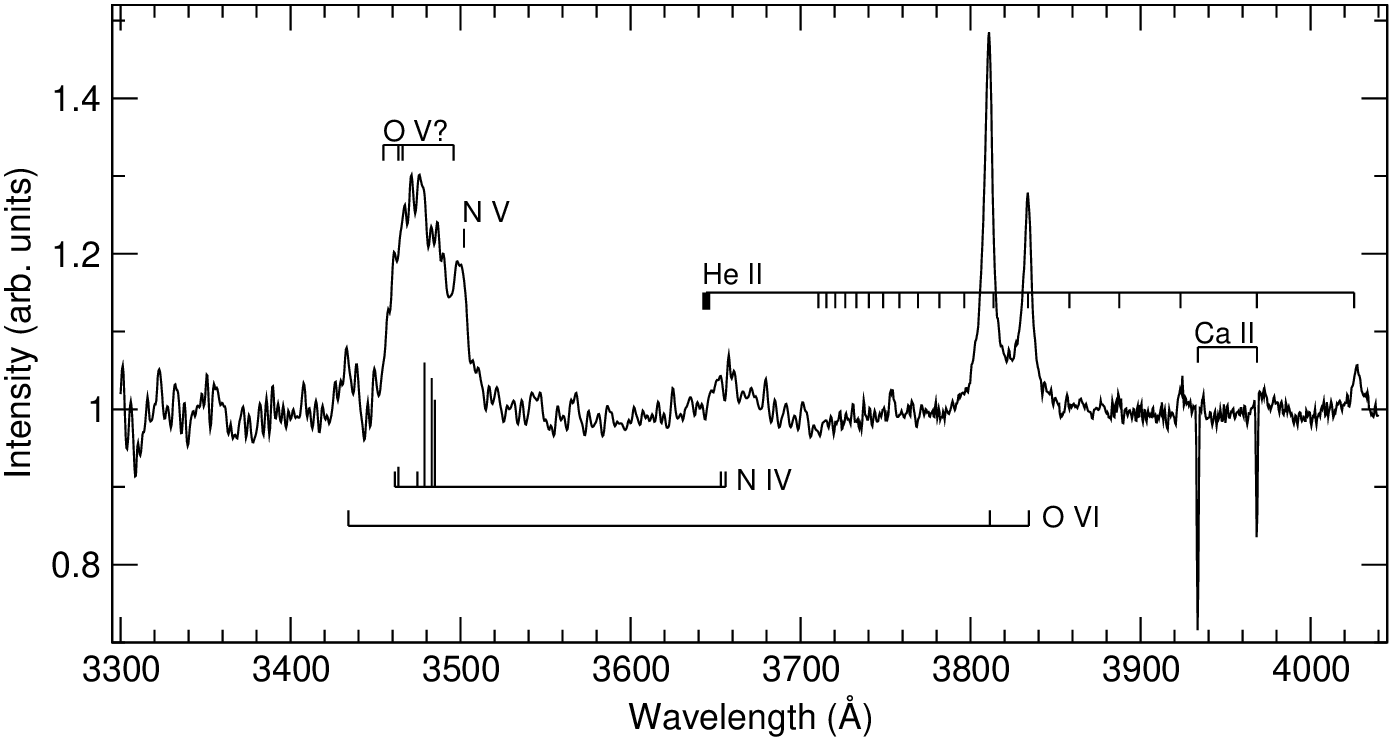}}
      \caption{The average blue spectrum of DI~Cru.  \label{spec2} }
\end{figure*}

We have compiled the CNO emission lines from the literature 
and organized them in a table with similar configurations, term by term (see Table~\ref{cno}). From 
this table it is clear that the spectrum is dominated by only three high ionization species: \ion{He}{2}, \ion{N}{5} and \ion{O}{6}. The only 
\ion{O}{5} -- \ion{N}{4} -- \ion{C}{3} lines that have been observed are the terms ($2p~^{1}P^{0}~-~ 2p^{2}~^{1}D$) and 
($3s~^{3}S~-~3p~^{3}P^{0}$).

We used Coud\'e spectra centered at 8300{\AA} to search for spectroscopic evidence of a possible secondary star. 
These spectra were divided by the spectra of hot calibration stars to eliminate the atmospheric absorption lines. 
We also obtained spectra of cool comparison stars (of M, G and K spectral types) with the same instrumental configuration and did the same 
process in order to compare the characteristic spectral lines of the cool stars to the lines of DI~Cru. We could not 
see any evidence of a cool component. In our high-resolution FEROS spectra we were also unable to detect any 
spectral feature of a hypothetical secondary star.

The terminal wind velocity, as estimated from \ion{He}{2} emission lines, is $v^{\star}$(HWZI)$= 2200$ km s$^{-1}$.

\placefigure{spec3}
\begin{figure*}
      \resizebox{\hsize}{!}{\plotone{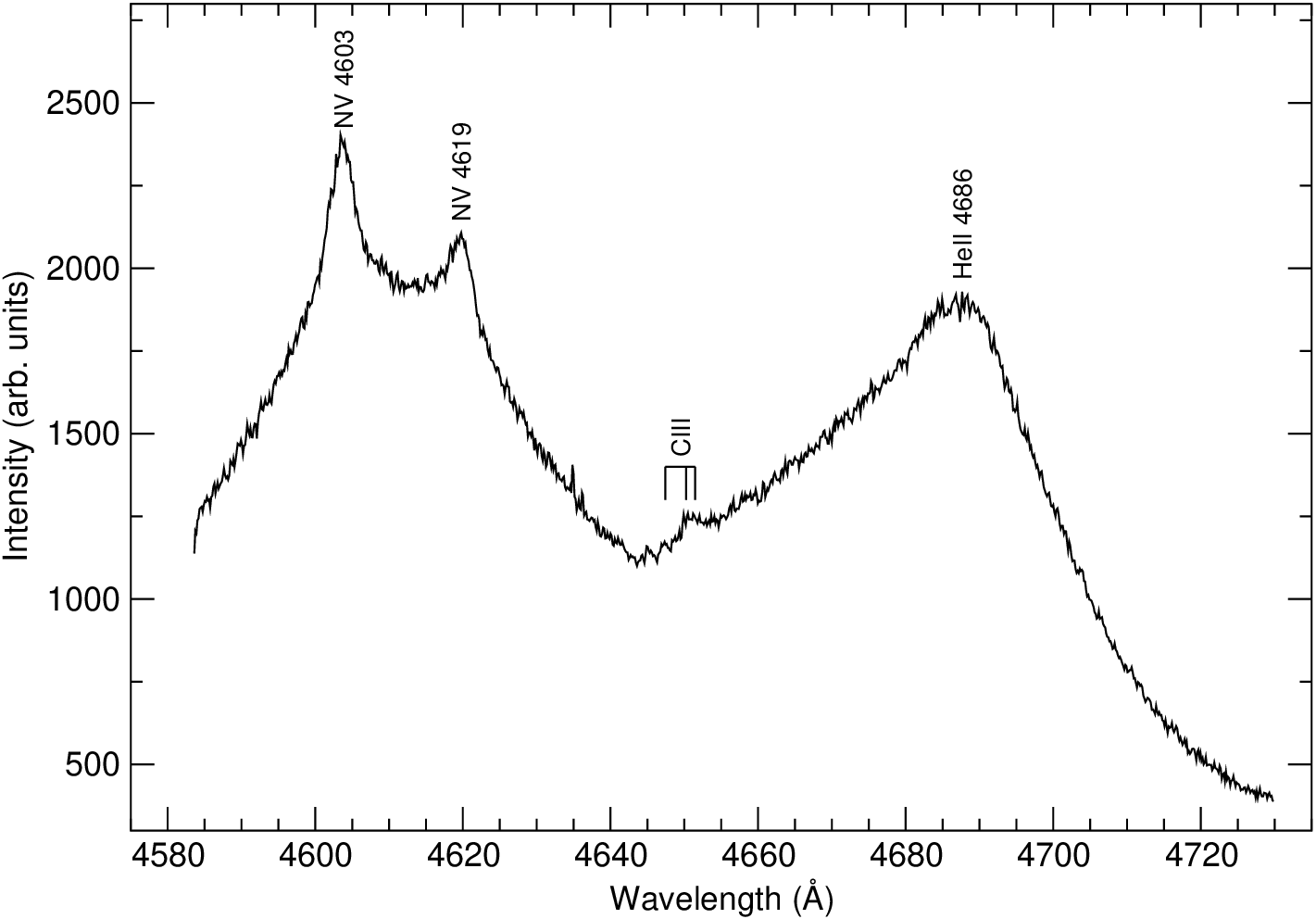}}
      \caption{The average of the five high resolution spectra obtained on 1999 April 1, centered at 4650 {\AA}.
      FWHM resolution is 0.23{\AA}.  \label{spec3} }
\end{figure*}

\placetable{cno}

\subsection{ Line variability}

The line profile and intensity variability has been extensively discussed by Mar00 and by Vee02b. Line fluxes vary from long time-scales, 
correlated with the overall photometric variability, to time-scales as short as minutes, with flares in the He II lines.

The ratio between the \ion{N}{5} 4603/19{\AA} and \ion{He}{2} 4686{\AA} seems to be quite variable, showing significant variations from night to night. 
This variability is well illustrated by comparing the spectrum from 1999 April 1,  (Figure~\ref{spec3}) with the average spectrum shown in 
Figure~\ref{tvs}. On 1999 April 1 the \ion{N}{5} 4603/19{\AA} was unusually strong, being more intense than \ion{He}{2} 4686{\AA}.
 A significant variability in the \ion{N}{5} to \ion{He}{2} line ratio was also noticed by Vee02b.

In order to study the line variability in a more objective way, we computed the Temporal Variance Spectrum (TVS). In this procedure
the temporal variance is calculated for each wavelength pixel from the residuals of the continuum normalized spectra. In our 
TVS analysis we calculate the square root of the variance as a function of wavelength. For further details and discussion of this method, 
see \citet{fulle}. The TVS of our medium resolution LNA spectra (81 spectra obtained on 1998 April 10, 11 and 12) is shown in Figure~\ref{tvs}.
The first thing to notice is the striking difference between the TVS profile of the \ion{N}{5} 4945{\AA} (narrow) line and the (broad) 
lines of \ion{He}{2} and
\ion{N}{5} 4603/19{\AA}. This illustrates well the two regimes described in \citet{fulle}: the two peaked profile resulting from a radial 
velocity variation and a true line profile variation -- $lpv$. When the line has no intrinsic variation but is 
displaced by radial velocity only, it introduces a double peak profile as 
 illustrated in Figure 1 of \citet{fulle}. In this case one expects $\sigma /I \sim$ K/FWHM. In the case of \ion{N}{5} 4603{\AA} and 
 of \ion{He}{2} 4686{\AA}, the TVS profile is much more similar to a true line profile variation, where the TVS profile is filled and the two peaks disappear. 

 Another remarkable characteristic is the difference in $\sigma/I$ for the \ion{He}{2} 4686{\AA} ($\sigma/I \sim5\%
 $) 
  and \ion{N}{5} 4603{\AA} ($\sigma/I \sim10\%
  $). This indicates 
 that the observed profile of \ion{N}{5} 4603{\AA} varies more than \ion{He}{2} 4686{\AA} (by about a factor of two).

Finally, the \ion{He}{2} 4686{\AA} shows a $\sigma/I$ ratio that depends on the velocity along the profile.
While the line profile has a maximum red-shifted by $\Delta v = +175$~km~s$^{-1}$, the TVS has its maximum at negative velocity
 ($\Delta v = -350$ km~s$^{-1}$). We 
tentatively explain this effect in terms of a variable P Cyg absorption, causing both the displacement of
the emission peak to the red and maximum variance in the blue.

\subsection{Continuum fluorescence $vs.$ recombination}

One noteworthy characteristic of the spectrum of this star is that the line widths of the \ion{N}{5} lines at 4945{\AA} and at 4603/19{\AA} 
are very distinct. While the 4945{\AA} 
line has FWHM $\sim 455$ km~s$^{-1}$, the 4603/19{\AA} lines have a narrow and a strong broad component of width similar to that
 of \ion{He}{2} 4686{\AA}, that is, FWHM$\sim 2~000$ km~s$^{-1}$. \ion{N}{5} is, therefore, emitted in two different regions, with distinct kinematics. 
What does this mean in terms of physical interpretation?

The first point to notice is that the 4945{\AA} line is a blend of a set of numerous recombination lines from the (6~--~7) transitions and also from 
the (8~--~11) transitions. The doublet 4603/19{\AA} (transition $3s~^{2}S$~--~$3p~^{2}P^{0}$) is formed by recombination from the 
cascading down process 
and also from the UV continuum fluorescence. The \ion{O}{6} 3811/34{\AA} lines have similar structure. 

The recombination emission depends on the square of 
the density while the continuum fluorescence lines have a linear dependence with density. 
If one assumes a wind that is optically thin for the exciting photons responsible for the fluorescence process and a velocity field given by

\begin{equation}
v(r) = v^{\star}\left(1- \frac{r^{\star}}{r}\right)^{\beta}				
\end{equation}
	
where $v^{\star}$ is the terminal velocity of the wind, $r^{\star}$ is the radius of the star and $\beta$ is a measure
of the rate of acceleration \citep{cast}, then the ratio of emissivity due to 
fluorescence ($E_{fl}$) and to recombination ($E_{rec}$) is

\begin{equation}
\frac{E_{fl}}{E_{rec}} \propto \left(1-\frac{r^{\star}}{r}\right)^{\beta}
\end{equation}

This shows that the ratio $E_{fl}/E_{rec}$ increases with radius to its maximum at $v=v^{\star}$. The narrower recombination lines 
suggest that they are preferentially emitted in the inner part of the wind where density is higher and velocity is smaller. 
The continuum fluorescence lines are preferentially produced at higher velocities in the wind -- similar to the ones 
seen in the \ion{He}{2} emission.

There may be a hint about the nature of the \ion{N}{5} 4603{\AA} line variability in the line profile calculated by CSH95. In their Figure 6, the 
line profiles from the two models considered are quite distinct. The two models have different effective temperatures and mass 
losses. The recombination lines are more sensitive to density and, therefore, to mass loss. 

What we see in our data is that the 
recombination lines are less variable than the continuum fluorescence lines. The \ion{N}{5} 4945{\AA} recombination blend, for 
example, does not seem to vary as its TVS shows a velocity displacement pattern and not a $lpv$ type profile. The 
continuum fluorescence lines are very sensitive to the central source effective temperature. A variable central source temperature is, 
therefore, the likely explanation for the \ion{N}{5} 4603/19{\AA} variability. This variability in temperature could be originated either in the 
stellar atmosphere or, more likely, in the optically thick wind with variable optical thickness. A variable optical depth in the wind could,
in principle explain, at the 
same time, the high variability in the \ion{N}{5} lines, the photometric variability and the fact that the star becomes bluer when 
fainter (Vee02a). 
CSH95 have calculated detailed models for the emission line profiles. In their figure 6 they show that the two models with effective
temperatures of T$_{eff} = 80~000 $ K and T$_{eff} = 89~000$ K both describe reasonably well the line \ion{N}{5} 4945{\AA} but the lines 
\ion{N}{5} 4603/19{\AA} are only adjusted well if the temperature is about T$_{eff} = 89~000$ K. This again argues in the sense that 
the intensity of these lines is much more sensitive to the stellar temperature than the  \ion{N}{5} 4945{\AA} line.

  We see significant evidence of variability in the degree of ionization. For instance, although we can not identify
  the \ion{N}{4} multiplet near 7123{\AA} in the FEROS (2002) spectrum, in our spectrum taken in the 3400--3900{\AA} region
  (1996) there is a strong emission of the \ion{N}{4} 3479{\AA} multiplet. These two multiplets arise in the recombination cascading and
  correspond to successive transitions; being correlated, they should be both present or absent. An explanation for this 
  apparent discrepancy may be that the star was at a higher degree of ionization in January 2002, when compared to June 1996
  (see also a discussion about the variability of the \ion{N}{4} line in Vee02a).

  The degree of ionization of a stellar wind is related to the ionization parameter (the density of ionizing photons divided by
  the electron density).
  The correlation between the \ion{N}{5} 4603/19{\AA} intensity and the appearance of the \ion{C}{3} line in Figure~\ref{spec3}
  (when compared to Figure~\ref{tvs}) seems to point toward opposite directions. While a stronger 
  emission of the \ion{N}{5} lines in Figure~\ref{spec3} suggest a higher temperature, the presence of the low 
  ionization species \ion{C}{3} can only be reconciled if a high wind density is also assumed.

\section{The 0.3319 d period}      \label{orbper}

Radial velocity curves were constructed using \ion{N}{5} 4603{\AA} (spectra
with 0.65{\AA} and 0.23{\AA} spectral resolution), \ion{N}{5} 4945{\AA},
\ion{He}{2} 4686{\AA} and \ion{O}{6} 3811{\AA}  (0.65{\AA}
resolution) emission lines measured at the Coud\'e spectrograph.
\ion{N}{5} 4945{\AA}, \ion{N}{5} 4603{\AA} and \ion{O}{6} 3811{\AA} from the FEROS spectra (0.1{\AA} resolution) were also used.
 The determination of the radial 
velocity from the \ion{N}{5} lines was made by fitting a Gaussian profile to the peak of the lines. In the particular case of the \ion{N}{5} 4603{\AA}
 we adjusted a 
Gaussian using the profile above 80 \% of its height, given the proximity of the line at 4619{\AA} and also because of the broad (fluorescence) 
component that may have a somewhat distinct kinematics. The radial velocity of the \ion{He}{2} emission line was measured from the 
flux weighted centroid of the line. 

The \ion{O}{6} 3811{\AA} and the \ion{O}{6} 3834{\AA}
  lines are very close to each other. 
 Therefore the determination of their central wavelength was performed by deblending using two Gaussian components. 
 All radial velocities were corrected to heliocentric 
 velocity system and are given in Tables~\ref{vrtaba} and~\ref{vrtabb}.

\placetable{vrtaba}

\placetable{vrtabb}
 
 The various lines in consideration here show distinct systemic velocities (see Table~\ref{orbit}). This is so because of blending and self-absorption 
(in the case of \ion{He}{2}). All lines have, in fact blending problems. \ion{N}{5} 4603/19{\AA} are blended with 
each other and also with \ion{He}{2} 4686{\AA}. 
The broad and variable component leaves the situation even worse. In the case of the \ion{O}{6} lines, the situation is similar but somewhat 
better as it is not blended with any other strong line. It seems that the most reliable radial velocity measurements come from \ion{N}{5} 4945{\AA}. 
This is, in fact, a blend of a forest of \ion{N}{5} \mbox{6--7} (and also 8--11) transitions that happen to have similar wavelength. Since all are recombination 
lines and are emitted in the densest part of the wind, it seems to be a good indicator of the stellar kinematics. For this reason we subtracted
 from each radial velocity measurement of \ion{N}{5} 4603{\AA}, the systematic difference between the average velocity of this 
 line and that of \ion{N}{5} 4945{\AA}. 
 This difference is 46 km~s$^{-1}$. We also assigned different weights to spectra with distinct resolution: 
 weight 2 to spectra with resolution 0.65{\AA} and 
 weight 3 to spectra with resolution 0.23{\AA} and 0.10{\AA}.

 \placetable{orbit}

As found by \citet{marche1}, our data also show that DI~Cru presents photometric variations with time-scales of months. The total 
amplitude of variations we observed is of about 0.45 mag. This seasonal type of variation is unique among WR stars.  We also found that 
the photometric characteristics such as flickering activity and night to night light curve shape depend strongly on the level of the overall 
brightness. For this reason we separated the data in three groups, depending on the intensity. We will call these groups as minimum, 
intermediate and maximum brightness levels and will, as a first approach, analyze them separately in terms of 
periodicity searching routines. The minimum brightness level was arbitrarily defined by photometric measurements with 
$\Delta$mag(v-c1)$ > -0.65$, where $\Delta$mag(v-c1)  is the difference in magnitudes between the variable star (DI Cru) 
and the comparison star c1 ($V$=11.79, $\alpha$(J2000)=12:05:24.5, $\delta$(J2000)=$-62$:01:43.6), in the V band. 
In the intermediate state $-0.85 > \Delta$mag(v-c1)$ > -1.00$, and in the maximum state $-0.95 > \Delta$mag(v-c1)$ > -1.10$.

\placefigure{per}
\begin{figure*}
      \resizebox{\hsize}{!}{\plotone{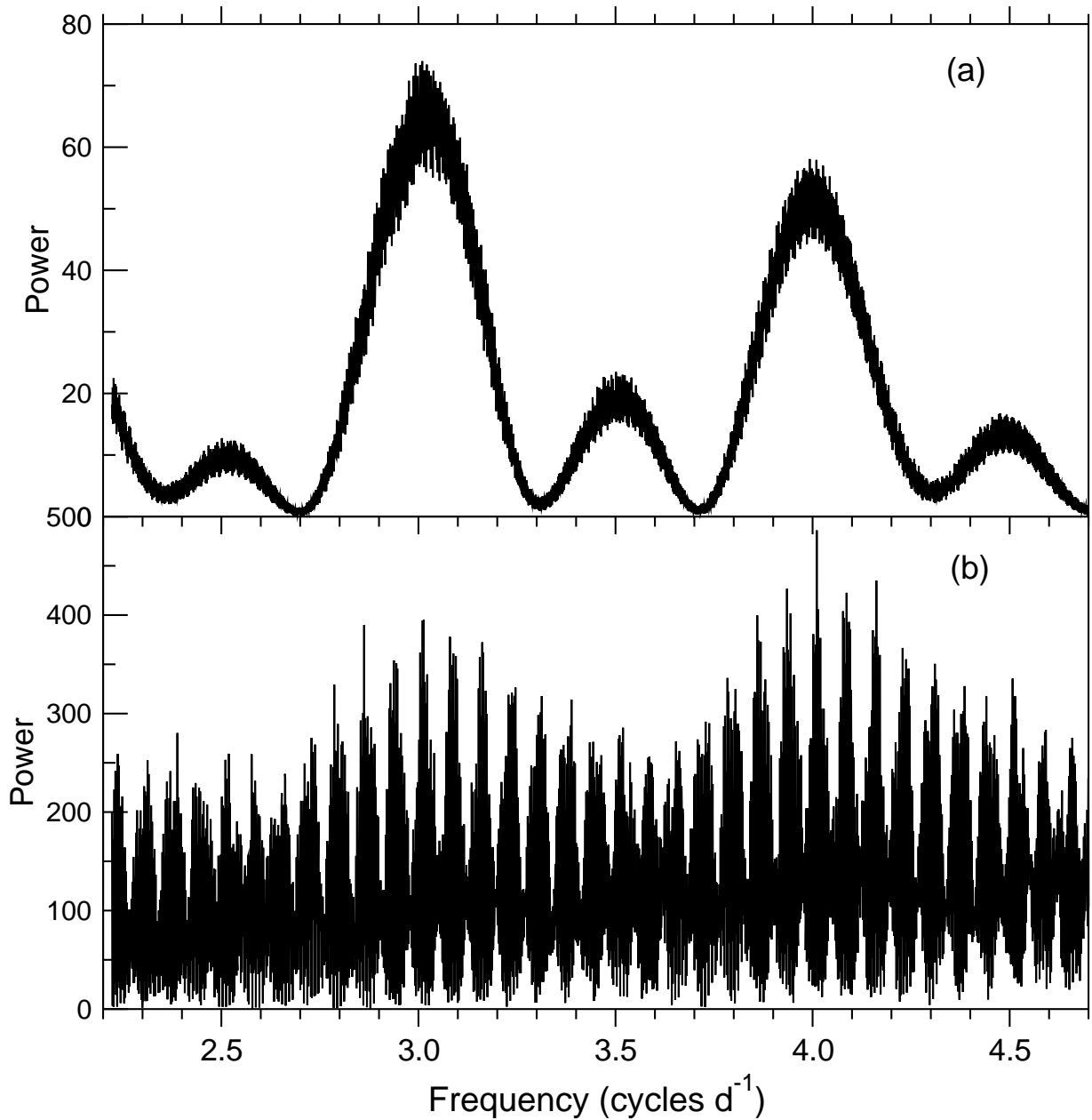}}
      \caption{Lomb-Scargle periodograms of the radial velocity (a) and intermediate brightness level photometric (b) data.\label{per} }
\end{figure*}

We initially applied the Lomb-Scargle algorithm to search for periodicities in the radial velocity and intermediate
level photometric data. 
The resultant periodograms are displayed in Figure~\ref{per}, and show a main signal with $P=0.3319$ d
 (3.013 cycles d$^{-1}$).
 An one day alias of 0.25 d (4.0 cycles d$^{-1}$) is also strong but less significant in the 
 radial velocity data. The one month aliases of 0.324 and 0.340 d (3.09 and 2.94 cycles d$^{-1}$)
 are also quite strong and can not 
 be discarded on the basis of these periodograms only. The high brightness level photometric data will be analyzed in Section~\ref{levels}.

The photometric ephemeris is:
\begin{equation}
T_{min} (HJD) = 2~450~917.297(\pm20) + 0.3319(\pm7) \times E
\end{equation}
where $T_{min}$ is the minimum of the intermediate level light curve (see Figure~\ref{lc1}).

The radial velocity curve is consistent with this period and the time of crossing from positive to negative values when compared to $\gamma$
is $T_{0} (HJD) = 2~450~917.232(\pm20)$.
Figure~\ref{vr} shows the radial velocity curve of the \ion{N}{5} lines folded with the 0.3319 d period. 
The best fit sine-wave is also shown and yields the values of $\gamma=-69(\pm 2)$ km~s$^{-1}$ for the systemic velocity and $K= 58 (\pm 2)$ km~s$^{-1}$ 
for the radial velocity semi-amplitude. We used the same values for the period and $T_{0}$ to plot the radial velocity curve of the \ion{He}{2} emission line. As can 
be seen in Figure~\ref{vrhe}, it presents a significant phase lag (0.2 in phase) when compared to \ion{N}{5}. This confirms the stratified nature of the 
wind as already shown by Vee02b.

\placefigure{vr}
\begin{figure*}
      \resizebox{\hsize}{!}{\plotone{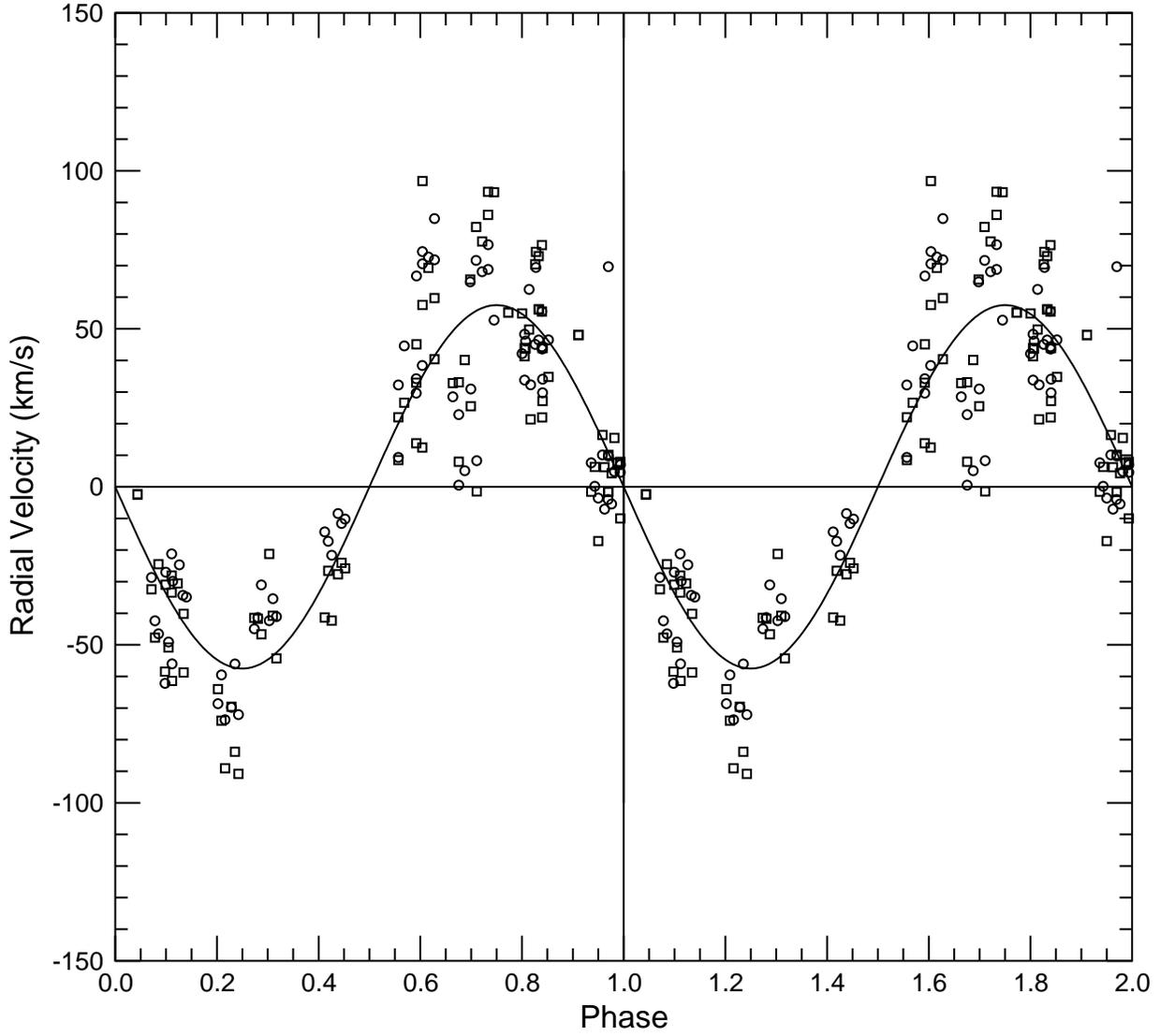}}
      \caption{Radial velocity curve of the  \ion{N}{5} 4603{\AA}  (squares) and
	\ion{N}{5} 4945{\AA} (circles) emission lines folded with $P=0.3319$ d and 
	$T_{0} = 2~450~917.232(\pm20)$ HJD.
	The systemic velocity has been subtracted from the data.  \label{vr} }
\end{figure*}

\placefigure{vrhe}

\begin{figure*}
      \resizebox{\hsize}{!}{\plotone{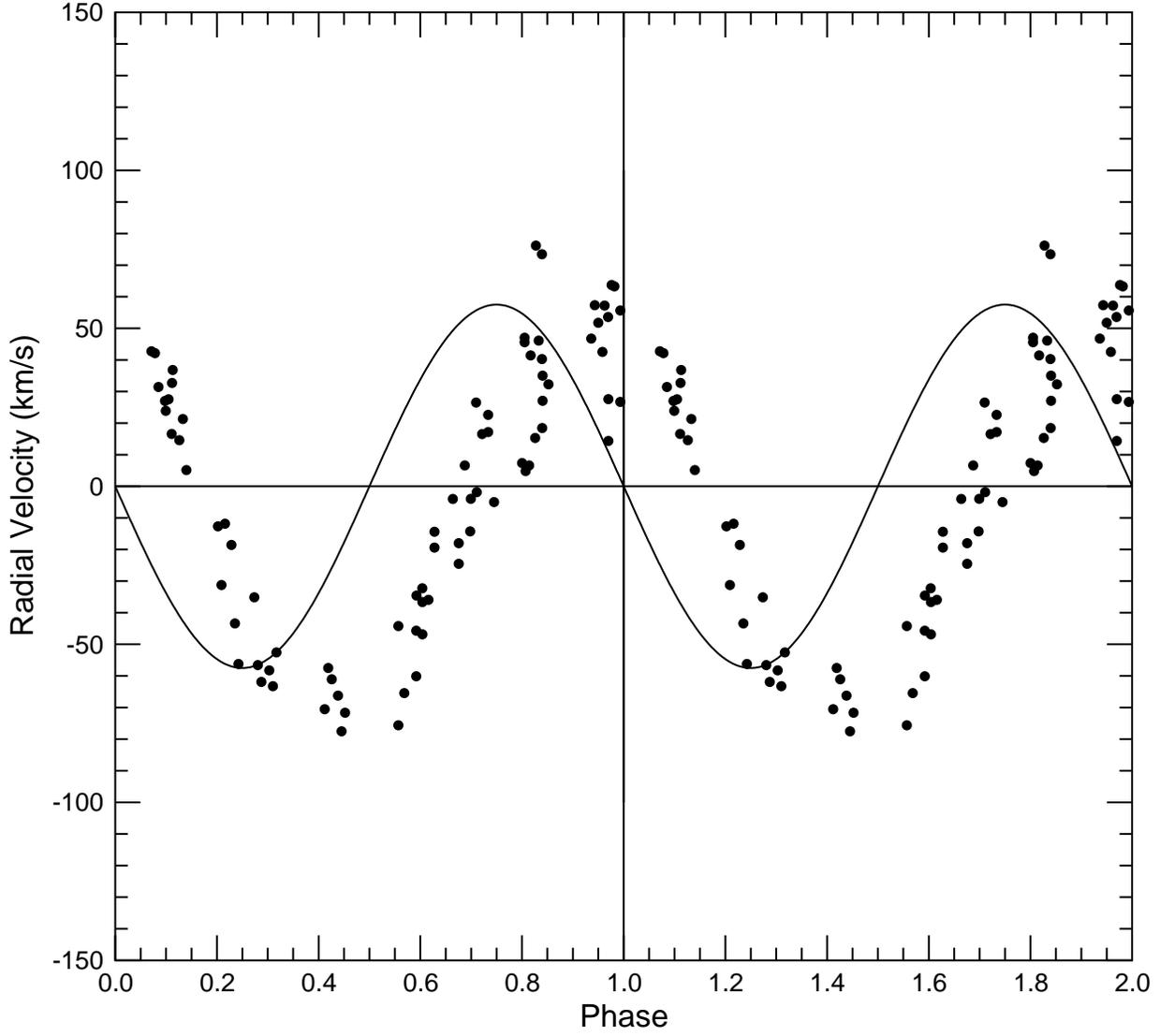}}
      \caption{ Radial velocity curve of the \ion{He}{2} emission lines folded with $P=0.3319$ d and 
	$T_{0} = 2~450~917.232(\pm20)$ HJD.
	The line represents the sinusoid fit to the \ion{N}{5}
	radial velocity data. The systemic velocity has been subtracted from the data. \label{vrhe}}
\end{figure*}

One should notice the strong difference between our values for $K$ and $\gamma$ and those measured by \citet{nie}. These 
authors found that the amplitudes of the radial velocity curves of the \ion{N}{5} 4603/19{\AA} and the \ion{He}{2} 4686{\AA} lines were,
respectively, 320 km~s$^{-1}$ and 173 km~s$^{-1}$, while the systemic velocity determined from these same lines were also quite discrepant, namely  $+4$
km~s$^{-1}$ and $+204$~km~s$^{-1}$.
 Similar differences between distinct lines have also been reported by 
Vee02b. This is, perhaps, not surprising as these lines are so badly blended. Spectral resolution also matters as the narrow component of 
\ion{N}{5} can only be measured with some confidence for medium to high resolution spectra.

The average light curve of the intermediate brightness level with the period of 0.3319 day is shown in Figure~\ref{lc1}. 
A double wave with two unequal minima is seen. 

\placefigure{lc1}
\begin{figure*}
      \resizebox{\hsize}{!}{\plotone{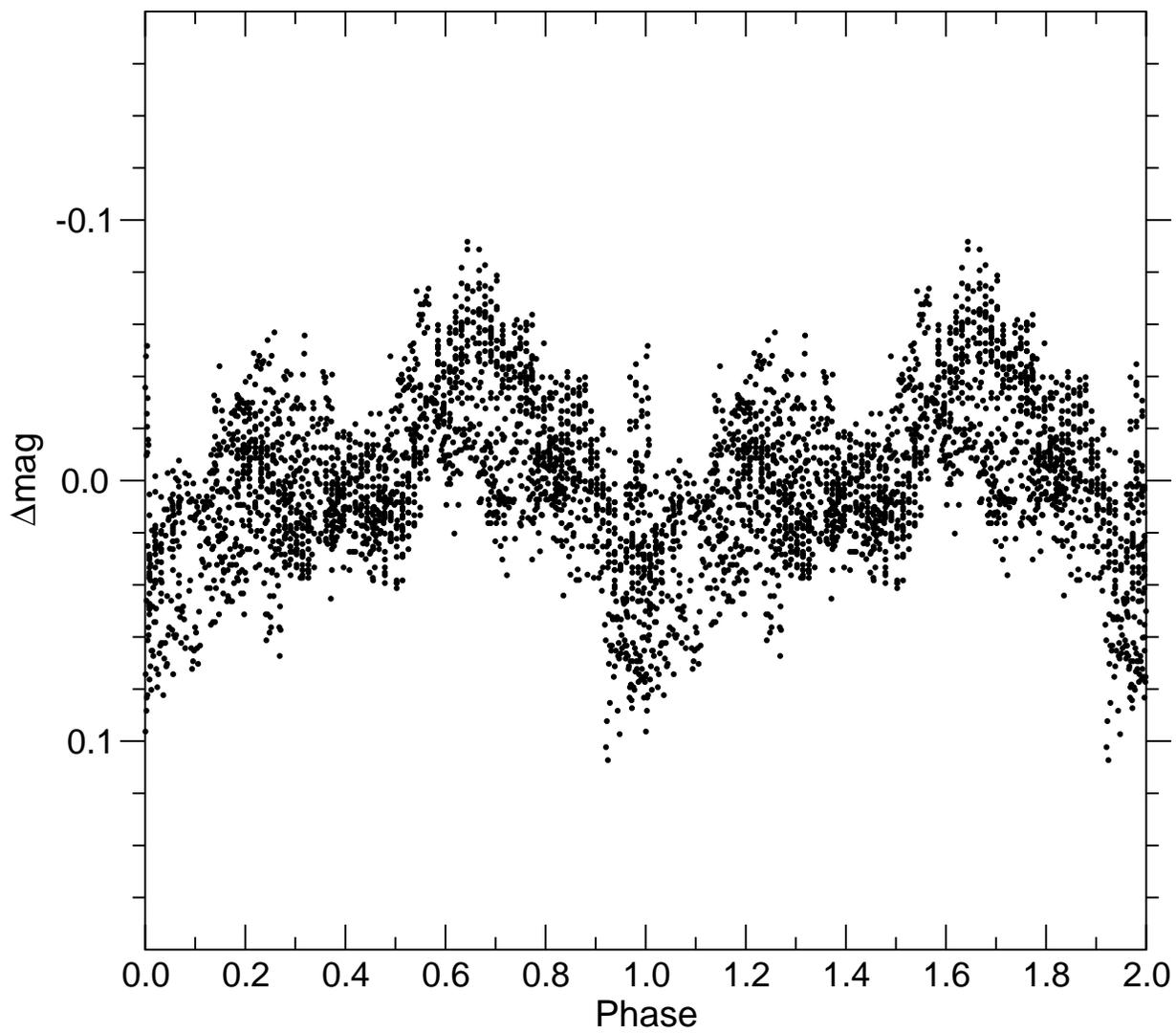}}
      \caption{ Intermediate brightness level average light curve of DI~Cru, plotted in phase using
	the photometric ephemeris. \label{lc1}}
\end{figure*}

\section{Multiple periodicities}\label{levels}

The behavior of the radial velocity and intermediate level photometric data seems to be consistent with a single period 
phenomenon -- the classical situation of a binary system. In the present analysis, however, the situation is clearly more complicated. 
This is well illustrated by the Lomb-Scargle periodograms for the photometric data obtained in the years 1997, 1998 and 
1999 (Figure~\ref{lombfot}). The 1997 data, in which the star was at the intermediate brightness level, 
is consistent with a single period of $P_s=0.3319$ d (3.013 cycles d$^{-1}$). In 1998 (intermediate/high levels) however, when 
radial velocity showed this same period, photometry shows a distinct value of about 0.154 d (6.49 cycles d$^{-1}$).
 
\placefigure{lombfot}
\begin{figure*}
      \resizebox{\hsize}{!}{\plotone{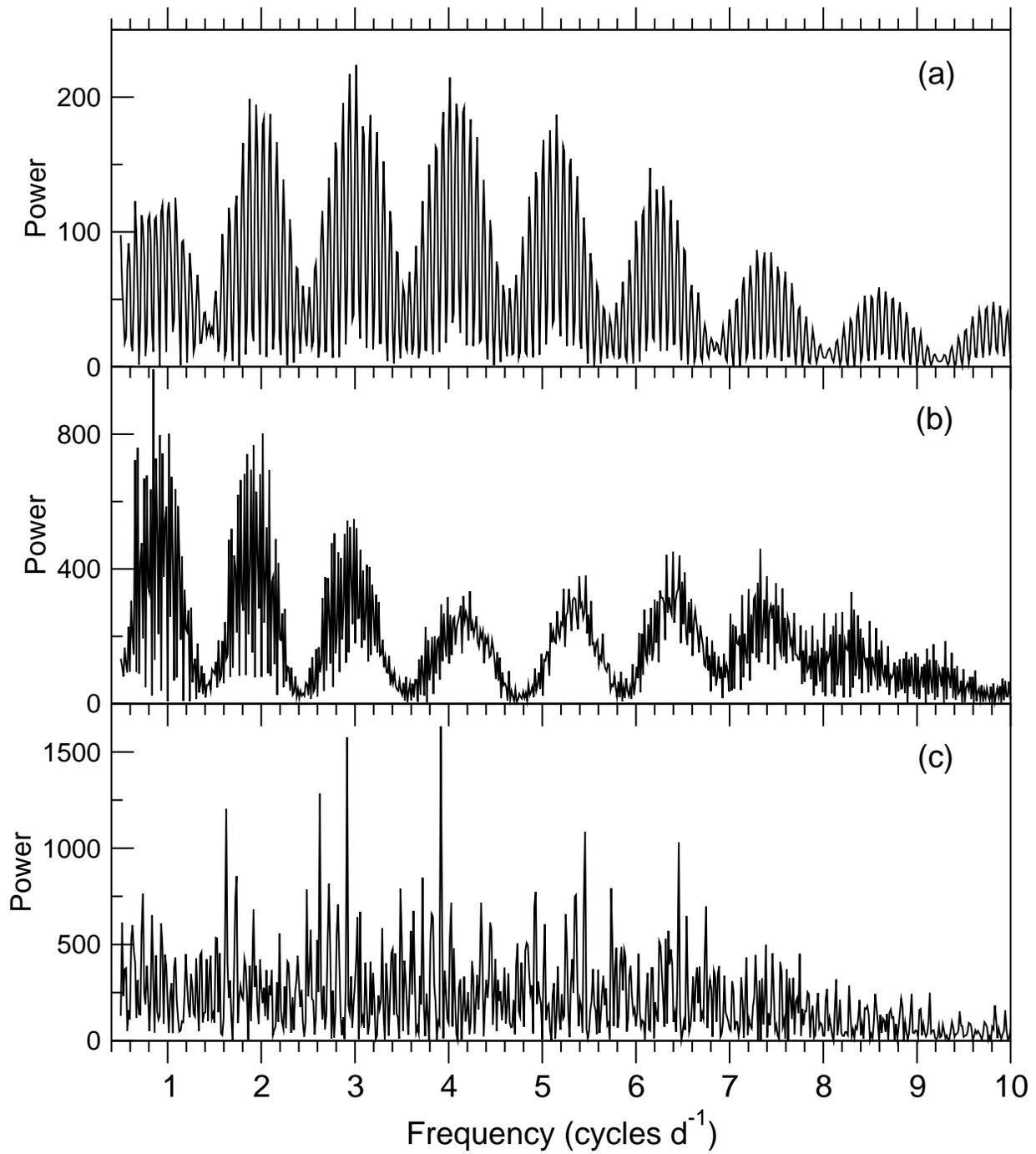}}
      \caption{Lomb-Scargle periodograms for the photometric data obtained in the years 1997 (a), 1998 (b) and 1999 (c). \label{lombfot} }
\end{figure*}

The situation becomes even more complicated in 1999. In this season we have the best coverage and the star was 
more variable, being at its maximum photometric level. At least 3 periods are seen: 0.378, 0.254 and 0.183 d (2.65, 3.94 and 5.46 cycles d$^{-1}$),
and none of them matches the previous values. An additional period of 0.583 d (1.72 cycles d$^{-1}$) seems to be present with weaker signal in 
the years of 1998 and 1999. Table~\ref{fre} summarizes the principal periods found for this object.

The radial velocity measurements made by \citet{marche2} -- Mar00 -- allowed those authors to determine a period of 0.329 d. 
As these observations were made in 1999 March and we have spectroscopic observations from 1999 April/May, it 
is worthwhile to combine them and refine the analysis of the possible periods. To do this, we incorporated to our set of 
1999 radial velocity data only the Mar00 data for the narrow \ion{N}{5} 4945{\AA} line, with weight 1 (our data kept weights 
as defined in Section~\ref{orbper}). The first thing to notice in the resultant periodogram (Figure~\ref{lomb99}) is that, 
as expected, the strongest signal is in the range of frequencies of 3.01 -- 3.05 cycles~d$^{-1}$ (0.328 -- 0.332 d). Surprisingly,
 there is a peak with about the same intensity at 3.74 cycles~d$^{-1}$ (0.267 d). Other two peaks of weaker intensity appear at 4.29 and 3.60 
 cycles~d$^{-1}$ (0.233 and 0.278 d). 
 These periods are very close to the photometric periods found in the 1989 and 1991 data by Vee02a. The values of 3.74  
and 3.60 cycles~d$^{-1}$ (0.267 and 0.278 d) are also close to the oscillations seen in radial velocity data in 1989/1991 (Vee02b).

 \placefigure{lomb99}
 \begin{figure*}
      \resizebox{\hsize}{!}{\plotone{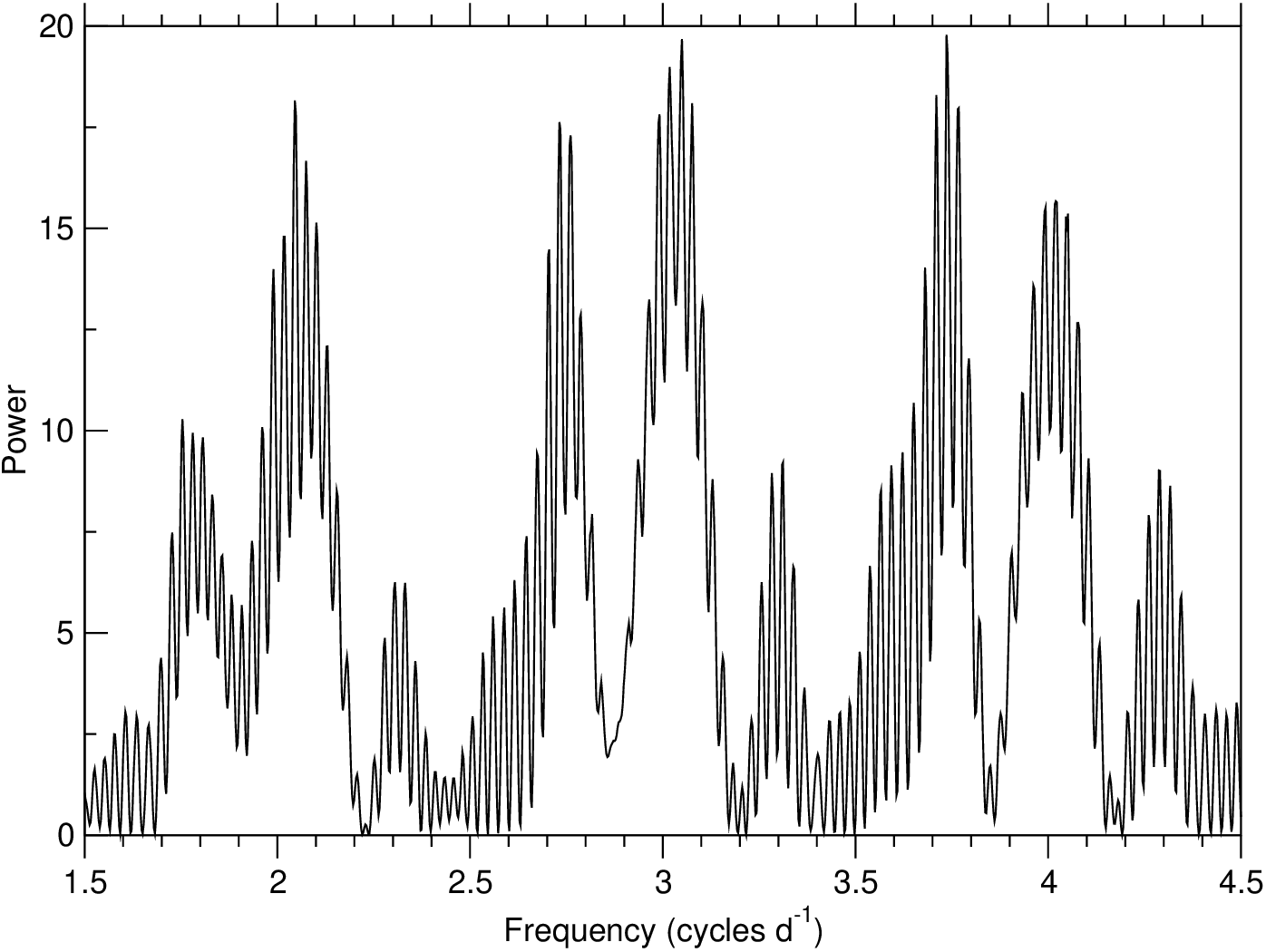}}
      \caption{Lomb-Scargle periodogram for 1999 (LNA plus Mar00) radial velocities
	for the \ion{N}{5} 4945{\AA} lines. \label{lomb99} }
\end{figure*}

Combining these data with our measurements from 1998 we obtain the strongest peak at 3.014 cycles~d$^{-1}$ (0.3318 d), which is about the same 
value (0.3319 d) determined in Section~\ref{orbper}.
In the same periodogram we also see a secondary peak at 3.55 cycles~d$^{-1}$ (0.282 d). These secondary radial velocity 
periods are very important in terms of interpreting the nature of the star. First, multiple periods have been seen in
 photometry before but not in radial velocity. Second, the secondary periodogram peaks seen in the 1998/1999 radial velocity data
 recover the photometric periods seen  in 1989/1991. This means that the star keeps its frequency memory and is not controlled by a 
 single clock with ever-growing period as suggested by Vee02b.

Mar00 and Vee02b observed a puzzling phenomenon in the radial velocity behavior of this star. Although periodicities 
with large amplitude are seen by both authors, from time-to-time they also observe a standstill so that the radial 
velocity does not vary at all along single nights. This unusual behavior seems to repeat with time-scales
of 2 -- 3 days.

Could this be caused by constructive/destructive interference between the distinct but simultaneously present
multiple periods? A close inspection of the patterns shown in Figure~\ref{lomb99} suggests that this might 
well be the case. As at least 4 periods seem to be present in the observations from 1999, one expects a rather 
complex pattern of interference.
We expect, however, a main beat period as a result of the interference of the two main periods (0.329 d and 0.267 d).
This gives a beat period of 1.43 d. But, given that this is about 3/2 of a day, the actual observations would 
have a recurrence time-scale of about 3.3 d. This is consistent with the observations of Mar00. This is a clear 
argument in favor of the explanation of the ''standstill puzzle'' as a consequence of interference between the periods
that are present simultaneously.

\section{Discussion}

One aspect that must be considered carefully in radial velocity measurements is the resolution of the spectra, as well as the 
line under consideration. For example, Mar00 measured the radial velocity of the line blend of 
\ion{He}{2}4686{\AA}+\ion{N}{5} 4603/19{\AA}. Given the 
high variability of the \ion{He}{2}/\ion{N}{5} line ratio, it is hard to tell what a radial velocity variation of such a blend would mean. A second
 point one must caution about is the mix of recombination and continuum fluorescence of the \ion{N}{5} and \ion{O}{6} 
 $3s~^{2}S$--$3p~^{2}P^{0}$ transition, 
 as they are emitted in distinct regions within the wind. The broad component, formed in the high velocity and 
 low density wind, probably due to fluorescence, may have a kinematics 
 that is quite distinct from that of the narrow component, due to recombination. This conclusion one derives from Figure~\ref{tvs}, considering
that the broad component of \ion{N}{5} has a kinematics similar to that of \ion{He}{2} 4686{\AA}. This last line must also be considered 
 with care. Variable P Cygni-type absorption introduces a radial velocity shift in the peak of the \ion{He}{2} lines that may mask or 
 substantially contaminate the radial velocity measurements. By comparing the average velocity of \ion{N}{5} 4945{\AA} to that of 
 \ion{He}{2} 4686{\AA}, we conclude that this effect amounts to $+130$ km~s$^{-1}$. From these considerations we believe that the most reliable 
measurements are those made from the peak of \ion{N}{5} and \ion{O}{6} lines as they are the likely tracers of the stellar kinematics.

\subsection{Is DI~Cru a binary system?}

\citet{nie} and Mar00, based on their radial velocity observations, have concluded that their spectroscopic period is likely to 
be the binary one. Although there is a difference between the two periods (0.311 d $vs.$ 0.329 d) one has to consider that the period determined 
by Mar00 and by us is consistent with the idea that our main radial velocity period may, perhaps, be stable and represent the orbital one. The 
existence of secondary spectroscopic frequencies (3.74, 3.55, 4.29, 3.60 cycles~d$^{-1}$, or 0.267, 0.282, 0.233, 0.278 d),
 however, shows how much care one has to take in analyzing this object.

The secondary spectroscopic periods found in 1998 and 1999 have recovered the photometric periods found in 1989/91. This strongly 
argues against the idea that the variability is controlled by a single clock with rapidly increasing clock-rate (Vee02a). On the 
contrary, it seems that distinct periods appear and reappear from time to time.

The existence of multiple spectroscopic periods also shows that the orbital period is not the only one. We conclude that 
we cannot rule out the hypothesis that DI Cru is a binary system; but its binary nature has not yet been proven. This could possible be done 
by demonstrating that the star shows long term coherence or by detecting the secondary star.

\subsection{Non-radial oscillations}

Vee02c have discussed in detail various possibilities to explain the observed characteristics of this star. Although not 
discarding the possibility of a binary system, they proposed non-radial oscillations as the likely cause of the observed 
photometric and spectroscopic variations. The main reason for this idea is that the photometric variations are not strictly 
periodic and multiple frequencies seem to be present at a given epoch.

We show a summary of the photometric and spectroscopic (radial velocity) periods identified so far in Table~\ref{fre}. Most of the 
periods are either close to 3.0 -- 3.7 cycles~d$^{-1}$ (0.270 -- 0.333 d) or to 6.5 -- 7.3 cycles~d$^{-1}$ (0.137 -- 0.154 d).
 As pointed out by \citet{gen}, if the photometric 
data that show $\sim 7$ cycles~d$^{-1}$ (0.143 d) are folded using twice the single-wave period, the light curve appear ellipsoidal with unequal minima. 
Moreover, the radial velocity data obtained simultaneously with the photometry are in better agreement with the double-wave than with the single-wave 
period. Therefore we show, in parenthesis in Table~\ref{fre}, half the frequency in those cases. The distribution of periods that appear both in 
spectroscopy and photometry is  now concentrated near 3.0 -- 3.2 (0.313 -- 0.333 d), 3.5 -- 3.7 (0.270 -- 0.286 d) 
and 4.3 cycles~d$^{-1}$ (0.233 d).

\placetable{fre}

Vee02c have convincingly suggested that the photometric and radial velocity variations may be interpreted as non-radial oscillations.
The presence of simultaneous multiple spectroscopic and photometric periods certainly requires such an interpretation, regardless of 
the binary nature of the star. The stratification of the wind, which is visible in the delay of the radial velocity curves (Figure~\ref{vrhe}) 
of \ion{He}{2} and of the narrow component of \ion{N}{5} ($\Delta\phi \sim 0.2$) is consistent with this interpretation.

The presence of multiple periods suggests that one may be dealing with multiple modes of oscillations, and we will
not attempt to identify these modes because of the temporal distribution of our data.

It is interesting to compare WR~46$\equiv$DI~Cru with HD~45166, which has been studied in detail by \citet{oliv}.
We have shown that HD~45166 has an orbital period of 1.59 d and additional oscillations with periods between 2.4 hr and 15 hr
that appear in distinct velocity ranges in the wind. These additional non-orbital periods are interpreted as non-radial oscillations 
of a helium main sequence star of mass $M_1=3.7 M_{\sun}$. 
The wind is optically thin given that one can observe atmospheric absorption lines from the hot star and this marks a significant 
difference between the two stars under consideration (DI~Cru and HD~45166). DI~Cru has an optically thick wind.
This difference likely results from the masses of the two stars which are extreme cases in the context of WNE objects.
As a consequence, their luminosities and mass-loss rates are very distinct, implying totally different wind
optical depths. It may be surprising that even so the stars present oscillations with similar periods. Within this context, we consider that 
DI~Cru may be regarded as a luminous counterpart of the qWR star HD~45166.

\subsection{Nature and classification} \label{natu}

Is this star a WR or a CBSS? 
Here we list several arguments in favor of the former scenario: a) One indication that the star is not a CBSS is its luminosity 
($log L/L_{\sun}~=~5.53$), as determined by CSH95, which is larger than the maximum Eddington 
luminosity for a white dwarf ($log L/L_{\sun}~=~4.81$). Other arguments in favor of a WR star are: b) evidence for non-radial oscillations with periods 
that seem to be compatible with the expected ones for helium burning stars; c) 
good agreement between the observed line profiles and the ones predicted by model calculations assuming that it is a 
population I WR star (CSH95) and d) mass-loss rate ($log \dot{M}(M_{\sun}~yr^{-1})~=~-5.2$)
 too large for the context of a CBSS. For a CBSS one expects a rate 100 times smaller. 

 The classification of DI~Cru in \citet{huc01} is WN3~pec while SSM96 propose WN3b~pec, where b stands 
for ''broad''.  These authors considered that the star does not have hydrogen. We showed in Section~\ref{hydr} that the star 
does have hydrogen and in a measurable amount. Therefore we propose a classification of WN3b(h)~pec
 in the SSM96 three dimensional classification system. This classification 
 shows some features that are unusual in the context of WR stars: objects classified as WN3 as well as WNb do not 
 usually show hydrogen. In addition, only three WN objects (WR~46, WR~48c and WR~109) in the catalog by \citet{huc01} 
show \ion{O}{6} emission. 

Although we interpret the object as a binary system with an orbital period of 0.3319 d, its true binary nature has not yet been 
clearly demonstrated. We suggest that further medium to high resolution spectroscopy of the purely recombination (and narrow) line 
\ion{N}{5} 4945{\AA} may reveal long term coherence. Similarly, additional photometry, preferentially when at medium photometric level,
 is also encouraged. 

\section{Conclusions}

	The main conclusions of this paper are:
\begin{enumerate}

\item The optical spectrum of DI~Cru is rich in strong emission lines of high ionization species, mostly dominated by \ion{He}{2}, \ion{N}{5}, \ion{N}{4}
and \ion{O}{6}.  Weak emission of \ion{C}{3} and H$\beta$ (and presumably of other Balmer lines) is also present. Emission lines 
have been compiled from the literature and identified from the ultraviolet to the infrared. In the UV, emission of \ion{O}{5} and \ion{N}{4} is 
also observed together with very weak emission of \ion{C}{4}.

\item The \ion{N}{5}/\ion{He}{2} line ratio varies by a significant amount from night to night. The TVS analysis shows that the \ion{He}{2} 4686{\AA}
 line has P Cyg-like variable absorption while \ion{N}{5} 4603/19{\AA} lines have a strong and broad variable component. We propose that this 
variability is due to continuum fluorescence caused by excitation from a source (stellar atmosphere/optically thick wind) of 
variable temperature. We also show that the object has variable degree of ionization, probably caused by wind density variation.

\item From our 1998/1999 radial velocity measurements we derived a main period  of 0.3319 d (3.013 cycles~d$^{-1}$ ) with an amplitude of 
$K=58$ km~s$^{-1}$. This period is similar to the one found by Mar00 from 1999 observations.

\item When at intermediate photometric level (1997), it is possible, also to derive a photometric period that is consistent with 0.3319 d.
Its light curve shows two minima of unequal depths. 

\item In the years of 1998/1999 multiple spectroscopic and photometric periods were present, including those reported 
for the years 1989/1991. This argues against the idea that the variability is controlled by a single clock with rapidly increasing
clock rate (Vee02a). 

\item Mar00 interpreted the period of 0.3319 d as the orbital one. We argue that the additional photometric and spectroscopic
periods are associated to non-radial pulsations, as proposed by Vee02c. 

\item We call attention to the similarity between the multiple oscillation periods present in the stars DI~Cru and HD~45166,
despite the extreme differences between these stars (in the context of WNE objects) due to their very distinct masses. Given the
 similarities, we suggest that it might be a luminous counterpart of the qWR star HD~45166.

\item Interference between the various 
 periods explains the disappearance of radial velocity modulation, seen from time to time. 

\item The most convincing evidence that the star is not a CBSS is its luminosity, as determined by 
CSH95, which is larger than the maximum Eddington luminosity for a white dwarf. Other arguments in favor of a WR star are:
 evidence for non-radial oscillation, good agreement in line profile fitting and large mass loss rate.

\item Although the object is a possible binary system with an orbital period of 0.3319 d, its true binary nature 
(long term coherence, detection of the secondary star) has not yet been
 demonstrated. We suggest that further medium to high resolution spectroscopy of the purely recombination (and narrow) line 
 \ion{N}{5} 4945{\AA} may perhaps reveal long term coherence. Similarly, additional photometry, preferentially when at medium photometric level, 
 is also encouraged.

\end{enumerate}

\acknowledgements

We would like to thank the referee, Dr. D. Gies, for his constructive comments
on a previous version of this paper.
Thanks are due to A. Bruch, D. Cieslinski, E. Oliveira, G. Quast and C. A. Torres for obtaining
photometry data and R. P. Campos for obtaining spectroscopy data. We are also grateful to 
F. J. Jablonski for kindly providing the DFT program used in the period search and to A. Kanaan for 
his detailed comments on the manuscript.

\newpage

\begin{deluxetable}{lcccllcl}
\tabletypesize{\footnotesize}
\tablecaption{Journal of photometric observations of DI~Cru.\label{jophoto}}
\tablewidth{0pt}
\tablehead{
\colhead{Date} & \colhead{Hours} & \colhead{Number} & \colhead{Exp. Time} & \colhead{Telescope} & \colhead{Detector} &
\colhead{Filter} & \colhead{Brightness} \\
\colhead{(UT)} & \colhead{of Obs.} & \colhead{of Exps.} & \colhead{(s)} & \colhead{} & \colhead{} &
\colhead{} & \colhead{Level} 
}
\startdata
1996 Apr 04 &  3.1     & 174    & 10 & B\&C 60cm   & CCD 048 & \textit{V}     &  Low         			 \\
1996 Apr 05 &  5.0     & 309    & 10 & B\&C 60cm   & CCD 048 & \textit{V}     &  Low       			 \\
1996 Apr 06 &  6.2     & 407    & 10 & B\&C 60cm   & CCD 048 & \textit{V}     &  Low         			 \\
1996 Apr 29 &  7.4     & 1098   & 10 & B\&C 60cm   & CCD 009 & \textit{V}     &  Intermediate      		 \\
1996 May 25 &  3.0     & 295    & 10 & Zeiss 60cm  & CCD 009 & \textit{V}     &  Intermediate      		 \\
1996 Jun 15 &  0.3     & 5      & 10 & Zeiss 60cm  & CCD 009 & \textit{V}     &  Intermediate      		 \\
1997 Mar 25 &  4.1     & 199    & 40 & Zeiss 60cm  & CCD 009 & \textit{V}     &  Intermediate   		 \\
1997 Mar 26 &  2.6     & 150    & 40 & Zeiss 60cm  & CCD 009 & \textit{V}     &  Intermediate   		 \\
1997 Apr 08 &  4.1     & 216    & 40 & Zeiss 60cm  & CCD 009 & \textit{V}     &  Intermediate      		 \\
1997 May 07 &  4.3     & 12373  & 1  & Zeiss 60cm  & CCD 301 & clear &  Unidentified  \tablenotemark{a} 	 \\
1997 May 10 &  3.6     & 11026  & 1  & Zeiss 60cm  & CCD 301 & clear &  Unidentified  \tablenotemark{a}     \\
1997 Jun 01 &  4.9     & 11224  & 1  & B\&C 60cm   & CCD 301 & clear &  Unidentified  \tablenotemark{a}    	 \\
1997 Jun 02 &  0.3     & 16     & 40 & B\&C 60cm   & CCD 301 & \textit{V}     &  Intermediate      		 \\
1997 Jul 10 &  2.9     & 7837   & 1  & B\&C 60cm   & CCD 301 & clear &  Unidentified  \tablenotemark{a}    	 \\
1998 Mar 03 &  2.0     & 68     & 40 & Zeiss 60cm  & CCD 048 & \textit{V}     &  Intermediate      		 \\
1998 Mar 04 &  0.7     & 31     & 40 & Zeiss 60cm  & CCD 048 & \textit{V}     &  Intermediate      		 \\
1998 Mar 05 &  4.1     & 140    & 40 & Zeiss 60cm  & CCD 048 & \textit{V}     &  Intermediate      		 \\
1998 Mar 06 &  1.3     & 331    & 40 & Zeiss 60cm  & CCD 048 & \textit{V}     &  Intermediate      		 \\
1998 Mar 16 &  0.2     & 11     & 20 & B\&C 60cm   & CCD 301 & \textit{V}     &  Intermediate      		 \\
1998 Mar 26 &  1.2     & 300    & 10 & Zeiss 60cm  & CCD 301 & \textit{V}     &  Unidentified \tablenotemark{a}  \\
1998 Apr 05 &  0.8     & 26     & 10 & B\&C 60cm   & CCD 301 & \textit{V}     &  Unidentified   \tablenotemark{a}  \\
1998 Apr 06 &  1.2     & 82     & 10 & B\&C 60cm   & CCD 301 & \textit{V}     &  High    			\\
1998 Apr 08 &  1.1     & 66     & 15 & B\&C 60cm   & CCD 301 & \textit{V}     &  High    			\\
1998 Apr 11 &  6.1     & 2135   & 10 & B\&C 60cm   & CCD 301 & \textit{V}     &  High    			\\
1998 Apr 12 &  3.7     & 1281   & 10 & B\&C 60cm   & CCD 301 & \textit{V}     &  High    			\\
1998 May 11 &  2.5     & 270    & 15 & B\&C 60cm   & CCD 301 & \textit{V}     &  Unidentified   \tablenotemark{a}  \\
1998 May 12 &  3.2     & 217    & 10 & B\&C 60cm   & CCD 301 & \textit{V}     &  Unidentified   \tablenotemark{a}  \\
1998 May 13 &  2.3     & 89     & 15 & B\&C 60cm   & CCD 301 & \textit{V}     &  Unidentified   \tablenotemark{a}   \\
1999 Mar 18 &  5.3     & 374    & 40 & B\&C 60cm   & CCD 101 & \textit{V}     &  High     			\\
1999 Mar 19 &  9.1     & 800    & 30 & B\&C 60cm   & CCD 101 & \textit{V}     &  High     			\\
1999 Mar 21 &  1.1     & 67     & 40 & B\&C 60cm   & CCD 101 & \textit{V}     &  High     			\\
1999 Apr 02 &  7.8     & 1180   & 8  & Zeiss 60cm  & CCD 301 & \textit{V}     &  High     			\\
1999 Apr 04 &  1.9     & 375    & 8  & Zeiss 60cm  & CCD 301 & \textit{V}     &  High     			\\
1999 Apr 19 &  7.5     & 1400   & 10 & Zeiss 60cm  & CCD 101 & \textit{V}     &  Unidentified  \tablenotemark{a} \\
1999 Apr 22 &  0.7     & 150    & 10 & Zeiss 60cm  & CCD 101 & \textit{V}     &  Unidentified  \tablenotemark{a}  \\
1999 Apr 23 &  3.8     & 2376   & 6  & Zeiss 60cm  & CCD 301 & \textit{V}     &  High    			 \\
1999 Apr 29 &  3.3     & 494    & 10 & Zeiss 60cm  & CCD 301 & \textit{V}     &  High    			 \\
1999 Apr 30 &  4.1     & 702    & 10 & Zeiss 60cm  & CCD 301 & \textit{V}     &  High    			 \\
1999 May 02 &  6.2     & 1254   & 8  & Zeiss 60cm  & CCD 301 & \textit{V}     &  High    			 \\
1999 May 03 &  6.9     & 1450   & 8  & B\&C 60cm   & CCD 301 & \textit{V}     &  High    			 \\
1999 Jun 22 &  1.1     & 217    & 10 & Zeiss 60cm  & CCD 301 & \textit{V}     &  High    			 \\
1999 Jun 23 &  3.6     & 632    & 10 & Zeiss 60cm  & CCD 301 & \textit{V}     &  High    			 \\
1999 Jun 24 &  3.0     & 580    & 9  & B\&C 60cm   & CCD 301 & \textit{V}     &  High    			 \\
1999 Jun 25 &  2.4     & 582    & 10 & B\&C 60cm   & CCD 301 & \textit{V}     &  High    			 \\
\enddata
\tablenotetext{a}{Data obtained without the \textit{V} filter or falling between the intermediate and high brightness levels.}
\end{deluxetable}

\begin{deluxetable}{lcccccll}
\tabletypesize{\footnotesize}
\tablecaption{Journal of spectroscopic observations of DI~Cru. \label{jospect}}
\tablewidth{0pt}
\tablehead{
\colhead{Date} &  \colhead{Instrument} &  \colhead{Grating} &  \colhead{Number} &
\colhead{Exp. Time} & \colhead{FWHM Res.} & \colhead{Coverage} & \colhead{Brightness} \\
\colhead{(UT)} &  \colhead{} &  \colhead{(l/mm)} &  \colhead{of Exps.} &
\colhead{(s)} & \colhead{({\AA})} & \colhead{({\AA})} & \colhead{Level}
}

\startdata
1996 Jun 10    &  Cassegrain & 1200       & 1   & 300    & 2    & 4175 to 5130  & Intermediate \tablenotemark{a}\\
1996 Jun 11    &  Cassegrain & 1200       & 1   & 900    & 2    & 3300 to 3970  & Intermediate \tablenotemark{a}\\
1996 Jun 12    &  Cassegrain & 1200       & 1   & 300    & 2    & 4175 to 5130  & Intermediate \tablenotemark{a}\\
1998 Apr 10    &  Coud\'e    & 600        & 21  & 180    & 0.7  & 4520 to 4960  & High   \tablenotemark{a}\\
1998 Apr 11    &  Coud\'e    & 600        & 36  & 180    & 0.7  & 4520 to 4960  & High   \tablenotemark{b}\\
1998 Apr 12    &  Coud\'e    & 600        & 24  & 180    & 0.7  & 4520 to 4960  & High   \tablenotemark{b}\\
1999 Apr 01    &  Coud\'e    & 1800       & 5   & 1200   & 0.23 & 4580 to 4730  &High    \tablenotemark{a}\\
1999 Apr 29    &  Coud\'e    & 600        & 2   & 600    & 0.7  & 3740 to 4040  &High    \tablenotemark{b}\\
1999 Apr 30    &  Coud\'e    & 600        & 3   & 900    & 0.7  & 3740 to 4040  & High   \tablenotemark{b}\\
1999 May 01    &  Coud\'e    & 600        & 11  & 1200   & 0.7  & 3740 to 4040  & High   \tablenotemark{a}\\
1999 May 02    &  Coud\'e    & 600        & 13  & 1200   & 0.7  & 3740 to 4040  &High    \tablenotemark{b}\\
1999 Jul 01    &  Coud\'e    & 600        & 6   & 300    & 0.7  & 8077 to 8523  & Unidentified            \\
2002 Jan 24    & FEROS    & \nodata & 4   & 900    & 0.1  &  3600 to 9000  & Unidentified         \\
2002 Jan 25    & FEROS    & \nodata & 2   & 900    & 0.1  &  3600 to 9000  & Unidentified         \\
2002 Jan 26    & FEROS    & \nodata & 3   & 900    & 0.1  &  3600 to 9000  & Unidentified         \\
2002 Jan 27    & FEROS    & \nodata & 2   & 900    & 0.1  &  3600 to 9000  & Unidentified         \\

\enddata
\tablenotetext{a}{Estimated from quasi simultaneous (1 day apart) photometry.}
\tablenotetext{b}{Determined directly from simultaneous photometry.}

\end{deluxetable}

\begin{deluxetable}{llcccccllccccccc}
\tabletypesize{\scriptsize}
\tablecaption{Emission line identification and properties.  \label{lineprop}}
\tablewidth{0pt}
\tablehead{
\colhead{Species}			& \colhead{$\lambda_{obs}$}		& \colhead{RV(HC)(2002)}	&
\colhead{$-W_\lambda$ (2002) }	& \colhead{FWHM (2002) }			& \colhead{$-W_\lambda$ (96/98/99) }	&
\colhead{FWHM (96/98/99) }		\\
\colhead{ID}			& \colhead{ ({\AA}) }			& \colhead{ (km~s$^{-1}$) }		&
\colhead{({\AA})}			& \colhead{(km~s$^{-1}$)}			& \colhead{({\AA})}			& 
\colhead{(km~s$^{-1}$)}
}
\startdata
O VI (6-7)	& 3433.7		& \nodata & \nodata 	& \nodata 	& 1.2		& 1135		\\
N IV (3s-3p)	& 3474:		&  \nodata & \nodata 	& \nodata 	& 30:		& 3600:		\\
N V (7-10)	& 3500:		&\nodata & \nodata 	& \nodata 	& 2:		& 770:		\\
N IV (1s-1p) ?	& 3660:		& \nodata & \nodata 	& \nodata 	& 4:		& 5000:		\\
O VI (3s-3p)	& 3810.5		& -67	& 3.2		& 460		& 4		& 670		\\		
O VI (3s-3p)	& 3833.6		& -50	& 1.0		& 370		& 2		& 705		\\		
He II (4-13)	& 4027.3		& 127	& 0.5		& 450		& \nodata	& \nodata	\\
He II (4-12)	& 4103.2		& 239	& 1.0		& 660		& \nodata	& \nodata	\\
He II (4-10)+H$\gamma$& 4341.2	& \nodata & 0.8	& 600		& \nodata	& \nodata	\\
N V (7-9)	& 4518.3		&  -8	& 1.2		& 450		& \nodata	& \nodata	\\
He II (4-9)	& 4542.9		&  87	& 1.1		& 490		& 1.4		& 640		\\
N V (3s-3p)	& 4603.5/4619.9	&  -15/-5	& 49 		& 2600 		& 55		& 2140		\\
He II (3-4)	& 4688.2		&  160	& 82		& 2460		& 85		& 2040		\\		
He II (4-8)+H$\beta$	& 4860.8	& \nodata & 9 & \nodata	& 9		& \nodata	\\
N V (6-7)	& 4944.6		&  2	& 7.9		& 395		& 5		& 440		\\		
O VI (7-8)	& 5289.8		&  -68	& 1.0		& 550		& \nodata	& \nodata	\\
He II (4-7)	& 5413.9		&  \nodata & 19		& \nodata	& \nodata	& \nodata	\\
N V (9-13)	& 5668:		&  \nodata &  0.6:		& 840		& \nodata	& \nodata	\\
He II (5-15)	& 6407		& 29	& 0.5		& 425		& \nodata	& \nodata	\\
N V (8-10)	& 6477.6		&  -48:	&  1.8		& 425		& \nodata	& \nodata	\\
He II (4-6)+H$\alpha$	& 6562.6	& \nodata & 46		& \nodata	& \nodata	& \nodata	\\
He II (5-13)	& 6682.6		&  -27	& 1.9		& 780		& \nodata	& \nodata	\\
N V (9-12)	& 6744.7		& -121	& 0.8		& 540		& \nodata	& \nodata	\\
O VI (8-9)	& 7714:		&  67	& 1.4:		& 430		& \nodata	& \nodata	\\
\enddata 
\tablecomments{2002: FEROS data (0.1{\AA} resolution); 1996: Cassegrain data (2{\AA} resolution);
		1998/1999: Coud\'e data (0.7{\AA} resolution)}                                                                               
\end{deluxetable}

\begin{deluxetable}{lccccc}
\tablecaption{The CNO emission line table for DI~Cru.  \label{cno}}
\tablewidth{0pt}
\tablehead{
\colhead{Transition}	& \colhead{O}	&	& \colhead{N}	&	& \colhead{C}	
}
\startdata
	& \bf{O V}	&	& \bf{N IV}	&	& \bf{C III}	\\
IP (eV)	& 113.90		&	& 77.47		&	& 47.89		\\ 
[5pt]
\textbf{Term}	& {\boldmath $\lambda$}\textbf{({\AA})} -- {\boldmath $W_{\lambda}$}\textbf{({\AA})} &	&
{\boldmath $\lambda$}\textbf{({\AA})} -- {\boldmath $W_{\lambda}$}\textbf{({\AA})} &	&
{\boldmath $\lambda$}\textbf{({\AA})} -- {\boldmath $W_{\lambda}$}\textbf{({\AA})}	 \\
$2p~^{1}P^{0}$~--~$2p^{2}~^{1}D$	& 1371  \hspace{\stretch{1}}  y	&	& 1719\hspace{\stretch{1}}y	&	&
2297\hspace{\stretch{1}} bl \\
[3pt]
$3s~^{3}S$~--~$3p~^{3}P^{0}$	& 2781  \hspace{\stretch{1}}y	&	& 3479\hspace{\stretch{1}}-30:	&	&
4647\hspace{\stretch{1}}wk \\
[10pt]
	& \bf{O VI}	&	& \bf{N V}	&	& \bf{C IV}	\\
IP (eV)	& 138.12		&	& 97.89		&	& 64.49		\\ 
[5pt]
\textbf{Term}	& {\boldmath $\lambda$}\textbf{({\AA})} -- {\boldmath $W_{\lambda}$}\textbf{({\AA})} &	&
{\boldmath $\lambda$}\textbf{({\AA})} -- {\boldmath $W_{\lambda}$}\textbf{({\AA})} &	&
{\boldmath $\lambda$}\textbf{({\AA})} -- {\boldmath $W_{\lambda}$}\textbf{({\AA})} \\
$2s~^{2}S$~--~$2p~^{2}P^{0}$	& 1032\hspace{\stretch{1}}\nodata		&	& 1239\hspace{\stretch{1}}-11.6: 	&	&
1548\hspace{\stretch{1}}wk	\\
[2pt]
	& 1038\hspace{\stretch{1}}\nodata	&	& 1243\hspace{\stretch{1}}\nodata	&	& 1551\hspace{\stretch{1}}\nodata	\\
[2pt]
$3s~^{2}S$~--~$3p~^{2}P^{0}$	& 3811\hspace{\stretch{1}}-3.2 &	& 4604\hspace{\stretch{1}}-39:	&	&
5801\hspace{\stretch{1}}$<0.15$ \\
[2pt]
	& 3834\hspace{\stretch{1}}-1.0 &	& 4620\hspace{\stretch{1}}-19: &	& 5812\hspace{\stretch{1}}np \\
[2pt]
$4s~^{2}S$~--~$4p~^{2}P^{0}$	& 9342\hspace{\stretch{1}}y	&	& 11331\hspace{\stretch{1}}\nodata	&	&
14335\hspace{\stretch{1}}\nodata \\
[2pt]
	& 9398\hspace{\stretch{1}}y &	& 11374\hspace{\stretch{1}}\nodata &	& 14362\hspace{\stretch{1}}\nodata \\
[2pt]
$5s~^{2}S$~--~$5p~^{2}P^{0}$	&18550\hspace{\stretch{1}}\nodata	&	& 22572\hspace{\stretch{1}}\nodata	&	&
28617\hspace{\stretch{1}}\nodata \\
[2pt]
	&19663:\hspace{\stretch{1}}\nodata	&	& 22654\hspace{\stretch{1}}\nodata &	& 28675\hspace{\stretch{1}} \nodata\\
[2pt]
(4-5)	& 1126	\hspace{\stretch{1}}\nodata	&	& 1620/55 \hspace{\stretch{1}}bl	&	& 2530	\hspace{\stretch{1}}\nodata	\\
[2pt]
(5-6)	& 2070	\hspace{\stretch{1}}\nodata	&	& 2981	  \hspace{\stretch{1}}y	&	& 4658	\hspace{\stretch{1}}bl	\\
[2pt]
(5-7)	& 1292	\hspace{\stretch{1}}\nodata	&	& 1860	  \hspace{\stretch{1}}\nodata	&	& 2907	\hspace{\stretch{1}}\nodata	\\
[2pt]
(6-7)	& 3435	\hspace{\stretch{1}}-1.2	&	& 4945	  \hspace{\stretch{1}}-7.9	&	& 7726	\hspace{\stretch{1}}tell	\\
[2pt]
(6-8)	& 2083	\hspace{\stretch{1}}\nodata	&	& 2998	  \hspace{\stretch{1}}y	&	& 4685	\hspace{\stretch{1}} bl	\\
[2pt]
(7-8)	& 5291	\hspace{\stretch{1}}-1.0	&	& 7618	  \hspace{\stretch{1}}y	&	& 11908	\hspace{\stretch{1}}\nodata	\\
[2pt]
(7-9)	& 3143	\hspace{\stretch{1}}\nodata	& 	& 4520	  \hspace{\stretch{1}}-1.2	&	& 7063	\hspace{\stretch{1}}np	\\
[2pt]
(8-9)	& 7715	\hspace{\stretch{1}}-1.4	&	& 11110	  \hspace{\stretch{1}}\nodata	&	& 17368	\hspace{\stretch{1}}\nodata	\\
[2pt]
(7-10)	& 2431	\hspace{\stretch{1}}\nodata	&	& 3502	  \hspace{\stretch{1}}-2:	&	& 5471	\hspace{\stretch{1}}np	\\
[2pt]
(8-10)	& 4494	\hspace{\stretch{1}}y?	&	& 6478	  \hspace{\stretch{1}}-1.8	&	& 10124	\hspace{\stretch{1}}\nodata	\\
[2pt]
(9-10)	& 11033	\hspace{\stretch{1}}np	&	& 15536	  \hspace{\stretch{1}}\nodata	&	& 24278	\hspace{\stretch{1}}\nodata	\\
[2pt]
(8-11)	& 3427	\hspace{\stretch{1}}\nodata	&	& 4943	  \hspace{\stretch{1}}y	&	& 7737	\hspace{\stretch{1}}np	\\
[2pt]
(9-11)	& 6202	\hspace{\stretch{1}}bl	&	& 8927	  \hspace{\stretch{1}}y	&	& 13954	\hspace{\stretch{1}}\nodata	\\
[2pt]
(10-11)	& 14590	\hspace{\stretch{1}}\nodata	&	& 21000	  \hspace{\stretch{1}}\nodata	&	& 32808	\hspace{\stretch{1}}\nodata	\\
[2pt]
(8-12)	& 2916	\hspace{\stretch{1}}\nodata	&	& 4198 	  \hspace{\stretch{1}}bl	&	& 6560	\hspace{\stretch{1}}\nodata	\\
[2pt]
(9-12)	& 4692	\hspace{\stretch{1}}bl	&	& 6747	  \hspace{\stretch{1}}-0.8	&	& 10543	\hspace{\stretch{1}}\nodata	\\
[2pt]
(10-12)	& 8284	\hspace{\stretch{1}}\nodata	&	& 11928 	  \hspace{\stretch{1}}\nodata	&	& 18635	\hspace{\stretch{1}}\nodata	\\
[2pt]
(11-12)	& 19180	\hspace{\stretch{1}}\nodata	&	& 27593	  \hspace{\stretch{1}}\nodata	&	& 43138	\hspace{\stretch{1}}\nodata	\\
[2pt]
(9-13)	& 3939:	\hspace{\stretch{1}}\nodata	&	& 5668	  \hspace{\stretch{1}}-0.6	&	& 8858	\hspace{\stretch{1}}\nodata	\\
[2pt]
(10-13)	& 6198:	\hspace{\stretch{1}}y?	&	& 8918:	  \hspace{\stretch{1}}bl	&	& 13946	\hspace{\stretch{1}}\nodata	
\enddata 
\tablecomments{Meaning of the codes: y = line present; bl = blended line; wk = weak line; np = line not present; tell = telluric line.}                      
\tablerefs{\footnotesize IR: \citet{schmu}; \citet{vre}. Optical: Vee02a; CSH95; \citet{vre}. UV: CSH95; \citet{vacca}. 
Rest wavelengths and line identifications were obtained from: NIST Atomic Spectra Database (http://www.physics.nist.gov/cgi-bin/AtData/main$_{-}$asd),
Atomic Line List (http://www.pa.uky.edu/$\sim$peter/atomic/) 
and Atomic Molecular and Optical Database Systems (http://amods.kaeri.re.kr/).		
 }                                                                    
\end{deluxetable}

 \begin{deluxetable}{lccc}
\tabletypesize{\tiny}
\tablecaption{Radial velocity: \ion{N}{5} and \ion{He}{2} data. \label{vrtaba}}
\tablewidth{0pt}
\tablenum{5a}
\tablehead{
\colhead{HJD}             & \colhead{RV(km~s$^{-1}$)  }
& \colhead{RV(km~s$^{-1}$)   } & \colhead{RV(km~s$^{-1}$)  }\\
\colhead{(2,400,000+)} & \colhead{N V 4603{\AA}   }
& \colhead{He II 4686{\AA}   }                      & \colhead{ N V 4945{\AA}   }
}
\startdata
50245.4794 &-59       &\nodata &-59 \\
50247.3950 & -19      &\nodata &-59 \\
50914.4293 &-13       & -5        & -62    \\
50914.4316 &1          & 27       & -40    \\
50914.4339 &5          & 6         & -27    \\
50914.4401 &-8         & 11       & -38    \\
50914.4424 &-8         & 25       & -42    \\
50914.4447 & -9        & 39       & -33    \\
50914.4783 &44        & 57       & -7    \\
50914.4806 &61        & 98       & -0    \\
50914.4829 &56        & 88       & -4    \\
50914.4870 &65        & 94       & 5    \\
50914.4893 &72        & 88       & -3    \\
50914.4916 &72        & 66       & -19    \\
50914.5204 &53        & 147     & -2    \\
50914.5227 &55        & 144     & -27    \\
50914.5250 &34        & 111     & -16    \\
50914.7737 &24        & 36       & -5    \\
50914.7760 &36        & 24       & -1    \\
50914.7783 &75        & 34       & 3    \\
50914.7824 &48        & 35       & 1    \\
50914.7847 &19        &  52      & 13    \\
50914.7870 &38        & 57       & 0    \\
50915.4619 &12        & 67       & -43    \\
50915.4642 &12        & 46       & -71    \\
50915.4665 &-13       & 53       & -49    \\
50915.4704 &19        & 78       & -67    \\
50915.4726 &4          & 67       & -41    \\
50915.4749 &-23       & 69       & -63    \\
50915.5071 &23        & 117     & -38    \\
50915.5093 &20        & 118     & -23    \\
50915.5116 &0          & 112     & -39    \\
50915.5177 &6          & 106     & -42    \\
50915.5201 &23        & 98       & -38    \\ 
50915.5224 &13        & 103     & -25    \\
50915.5583 &-5        & 114      & -62    \\
50915.5606 &-11      & 85        & -2    \\
50915.5629 &-23      & 99        & -62    \\
50915.5669 &-6        & 134      & -67    \\
50915.5692 &-13      & 127      & -67    \\
50915.5715 &-31      & 98        & -65    \\
50915.6054 &-52      & 95        & -99    \\
50915.6077 &-49      &  88       & -93    \\
50915.6100 &-55      & 108      & -102    \\
50915.6144 &-52      & 86        & -96    \\
50915.6167 &-80      & 92        & -106    \\
50915.6190 &-62      & 76        & -107    \\
50915.6679 &-68      & 9          & -103    \\
50915.6731 &-43      & 13        & -114    \\
50915.6754 &-62      & 8          & -107    \\
50915.6777 &-76      & 18        & -113    \\
50915.7092 &-63      & 0          & -86    \\
50915.7115 &-48      & 14        & -89    \\ 
50915.7138 &-64      & 10        & -93    \\
50915.7179 &-49      & 5          & -80    \\
50915.7202 &-45      & -7        & -83    \\
50915.7225 &-47      & -1        & -82    \\ 
50916.5019 &34       & 78        & -30    \\
50916.5042 &22       & 76        & -26    \\ 
50916.5065 &28       & 78        & -9    \\
50916.5104 &49       & 86        & -27    \\
50916.5127 &52       & 117      & -25    \\
50916.5150 &1         & 89        & -28    \\
50916.5470 &-23      & 118      & -64    \\
50916.5493 &-15      & 128      & -72    \\
50916.5516 &-39      & 123      & -75    \\
50916.5557 &-15      & 128      & -79    \\
50916.5580 &-23      & 125      & -76    \\
50916.5603 &-17      & 135      & -77    \\
50916.5918 &-54      & 114      & -100    \\
50916.5941 &-69      & 113      & -114    \\
50916.5964 &-46      & 102      & -118    \\
50916.6007 &-80      & 98        & -134    \\
50916.6030 &-72      & 99        & -121    \\
50916.6053 &-83      & 104      & -128    \\
50916.6353 &-85      & 58        &  -140    \\
50916.6376 &-95      & 40        & -131    \\ 
50916.6399 &-100    & 59        & -145    \\
50916.6440 &-91      & 52        & -142    \\
50916.6463 &-105    & 28        & -128    \\
50916.6486 &-112    & 15        & -144     \\
51270.6301 &34       &\nodata  &\nodata  \\
51270.6501 &35       &\nodata &\nodata  \\
51270.6760 &27       &\nodata &\nodata  \\
51270.7015 &-14      &\nodata &\nodata  \\
51270.7200 &-24      &\nodata &\nodata  \\
\enddata
\end{deluxetable}
 
\begin{deluxetable}{lcccc}
\tabletypesize{\tiny}
\tablecaption{Radial velocity: O VI, N V and He II data. \label{vrtabb}}
\tablewidth{0pt}
\tablenum{5b}
\tablehead{
\colhead{HJD}             & \colhead{RV(km~s$^{-1}$)      } & \colhead{RV(km~s$^{-1}$)        }
& \colhead{RV(km~s$^{-1}$)   } & \colhead{RV(km~s$^{-1}$)  }\\
\colhead{(2,400,000+)} & \colhead{ O VI 3811{\AA}     }                  & \colhead{ N V 4603{\AA}      }
& \colhead{  He II 4686{\AA}       }                      & \colhead{ N V 4945{\AA}         }
}
\startdata
50246.3905 & -83        &\nodata &\nodata &\nodata  \\
51298.5841 &  -64       & \nodata&\nodata&\nodata  \\
51298.5936 &  -85       &\nodata &\nodata&\nodata  \\
51299.5865 &  -70       &\nodata &\nodata&\nodata  \\
51299.6434 &  -105     &\nodata &\nodata&\nodata  \\
51299.6842 &  -101     &\nodata &\nodata&\nodata  \\
51300.4403 &  -116     &\nodata &\nodata&\nodata  \\
51300.4590 &  -104     &\nodata &\nodata&\nodata  \\
51300.4881 &  -69       &\nodata &\nodata&\nodata  \\
51300.5041 &  -78       &\nodata &\nodata&\nodata  \\
51300.5227 &  -25       &\nodata &\nodata&\nodata  \\
51300.5437 &  -42       &\nodata &\nodata&\nodata  \\
51300.5611 &  -6         &\nodata &\nodata&\nodata  \\
51300.5797 &  -3         &\nodata &\nodata&\nodata  \\
51300.5977 &  -6         &\nodata &\nodata&\nodata  \\
51300.6395 &  -37       &\nodata &\nodata&\nodata  \\
51300.6885 &  -140     &\nodata &\nodata&\nodata  \\
51301.4251 &  -34       &\nodata &\nodata&\nodata  \\
51301.4499 & -8          &\nodata &\nodata&\nodata  \\
51301.4660 &  -51       &\nodata &\nodata&\nodata  \\
51301.4820 &  -41       &\nodata &\nodata&\nodata  \\
51301.4982 & -26        &\nodata &\nodata&\nodata  \\
51301.5182 &  -75       &\nodata &\nodata&\nodata  \\
51301.5371 &  -74       &\nodata &\nodata&\nodata  \\
51301.5539 &  -93       &\nodata &\nodata&\nodata  \\
51301.5706 &  -103     &\nodata &\nodata&\nodata  \\
51301.5876 &  -102     &\nodata &\nodata&\nodata  \\
51301.6232 &  -98       &\nodata &\nodata&\nodata  \\
51301.6413 &  -123     &\nodata &\nodata&\nodata  \\
51301.6578 & -98        &\nodata &\nodata&\nodata  \\
52299.7142&-125&-39&203&-103\\
52299.7264&-161&-82&187&-124\\
52299.7686&-122&-80& 78 &-117\\
52299.8048&-90  &-71& 41  &-105\\
52300.7596&  5    & 65&220& 24   \\
52300.7881&  5    & 53&199& 16   \\
52301.7111&-33  & -10&135&-51  \\
52301.7234&-42  & -5  &182&-44  \\
52301.7799&-27  &  8  &224&-26  \\
52302.7112& 32   & 72&261& 17  \\
52302.7654& -6    & 18&280&-32  \\
\enddata
\end{deluxetable}

 \begin{deluxetable}{lll}
\tablecaption{Radial velocity curve parameters.\label{orbit}}
\tablenum{6}
\tablewidth{0pt}
\tablehead{
\colhead{Line} & \colhead{K       } & \colhead{$\gamma$  }\\
\colhead{}	&\colhead{(km~s$^{-1}$)  }	&\colhead{(km~s$^{-1}$)  }
}
\startdata
\ion{O}{6} 3811{\AA}  &  $28({\pm}9)$    &  $-76({\pm}7)$ \\
\ion{N}{5} 4603{\AA}  &  $59({\pm}3) $   &  $-24({\pm}2)$\\
\ion{N}{5} 4945{\AA}  &  $54({\pm}3) $   &  $-70({\pm}2)$ \\
\ion{He}{2}4686{\AA}  &  $61({\pm}2)$   &   $61({\pm}2)$  \\
\enddata
\end{deluxetable}

\begin{deluxetable}{clclc}
\tablecaption{Photometric and radial velocity frequencies. \label{fre}}
\tablenum{7}
\tablewidth{0pt}
\tablehead{
\colhead{Year} & \colhead{Photometry\tablenotemark{a} (cycles~d$^{-1}$)} & \colhead{Ref.\tablenotemark{b}} 
& \colhead{Radial Velocity (cycles~d$^{-1}$)} & \colhead{Ref.\tablenotemark{b}}
}
\startdata	
1989		& \textbf{7.08 (3.54)}; 4.34			& (1)		& 3.54				& (5)	\\
1990		& 7.06: (3.53:); 4.35:			& (1)                          &  \nodata                         		& 	\\
1991		& \textbf{7.34 (3.67)}; 3.59:			& (1)		& 3.67				& (5)	\\
1993/1994	& \nodata  				& 		& \textbf{3.21}			& (2)	\\
1995		& 3.7:					& (1)		& 3.7:				& (5)	\\
1997		& \textbf{3.01}				& (3)                          & \nodata               		                 & 	\\
1998		& \textbf{6.49 (3.24)}; 3.7:; 1.71:		& (1,3)		& \textbf{3.01}; 3.55		& (3)	\\
1999		& 2.65; \textbf{3.94}; 5.46 (2.73); 1.72:	& (3)		& \textbf{3.04}; 3.74; 4.29; 3.60	& (4, 3)	\\
\enddata
\tablenotetext{a}{Numbers in parenthesis are double-wave frequencies (half the photometric frequency -- see Vee02a); bold-face 
numbers are the dominant frequencies.}
\tablenotetext{b}{References: (1) Vee02a; (2) Niemela et al. (1995); (3) This paper; (4) Mar00; (5) Vee02b}
\end{deluxetable}

\end{document}